\documentclass{elsarticle}

\usepackage{latexsym,bm,amssymb,amsfonts,amsbsy,amsmath,graphics,graphicx,color}

\usepackage{multirow}%
\usepackage{amsthm}%
\usepackage{mathrsfs}%
\usepackage[title]{appendix}%
\usepackage{xcolor}%
\usepackage{textcomp}%
\usepackage{manyfoot}%
\usepackage{booktabs}%
\usepackage{algorithm}%
\usepackage{algorithmicx}%
\usepackage{algpseudocode}%
\usepackage{listings}%

\usepackage{geometry,graphicx,amssymb,amsfonts,amsmath,bm,url}
\usepackage{pifont}
\usepackage{url}
\usepackage{fancyhdr}
\usepackage{pstricks,pst-node,pst-text,pst-3d}

\newtheorem{theorem}{Theorem}

\theoremstyle{remark}

\def \RR {{\mathbb{R}}}

\def \x {{\bf x}}

\def \x {{\bf x}}

\DeclareMathOperator{\MSE}{\overline{{E^2}}}
\DeclareMathOperator{\MSEuw}{\overline{{E^2}}_{\mathrm{uw}}}
\DeclareMathOperator{\MSEcv}{\overline{{E^2}}_{\mathrm{cv}}}

\DeclareMathOperator{\diag}{diag}

\DeclareMathOperator{\vc}{vec}

\newrgbcolor{marrone}{0.6156 0.3164 0.0313}
\newrgbcolor{DBlue}{0 0 0.7}
\newrgbcolor{ts}{0.50 0.598 0.781}
\newrgbcolor{verde}{0 0.4 0}
\newrgbcolor{LBlue}{0.58 0.65 0.9}
\newrgbcolor{Blue}{0.1 0.00 0.60}
\newrgbcolor{rosso}{0.9633    0.1641    0.0039}

\newrgbcolor{giorgia}{0.1 0.00 0.60}
\newrgbcolor{felice}{0.9633    0.1641    0.0039}
\newrgbcolor{pierluigi}{0.50 0.598 0.781}

\def \us {\overline{u}}
\def \vs {\overline{v}}
\def \ts {\overline{t}}

\begin{document}
\begin{frontmatter}

\title{On the use of the principle of maximum entropy to improve \\ the  robustness of  bivariate spline least-squares approximation}

\author[bari]{P.~Amodio}
\ead{ pierluigi.amodio@uniba.it }
\author[firenze]{L.~Brugnano}
\ead{ luigi.brugnano@uniba.it }
\author[bari]{F.~Iavernaro\corref{cor1}}
\ead{ felice.iavernaro@uniba.it }
\cortext[cor1]{Corresponding author}

\address[bari]{Dipartimento di Matematica, Universit\`a degli Studi di Bari Aldo Moro, Italy}
\address[firenze]{Dipartimento di Matematica e Informatica ``U. Dini'', Universit\`a di Firenze, Italy}

\begin{abstract}
We consider fitting a bivariate spline regression model to data using a weighted least-squares cost function, with weights that sum to one to form a discrete probability distribution. By applying the principle of maximum entropy, the weight distribution is determined by maximizing the associated entropy function. This approach, previously applied successfully to polynomials and  spline curves, enhances the robustness of the regression model by automatically detecting and down-weighting anomalous data during the fitting process. To demonstrate the effectiveness of the method, we present applications to two image processing problems and further illustrate its potential through two synthetic examples. 

Unlike the standard ordinary least-squares method, the maximum entropy formulation leads to a nonlinear algebraic system whose solvability requires careful theoretical analysis. We provide preliminary results in this direction and discuss the computational implications of solving the associated constrained optimization problem, which calls for dedicated iterative algorithms. These aspects suggest natural directions for further research on both the theoretical and algorithmic fronts.
\end{abstract}


\end{frontmatter}

\section{Introduction}
\label{sec_intro}
Spline approximation is a widely used technique in numerical analysis, data fitting, and geometric modeling. Its ability to provide smooth and flexible representations of complex functions makes it highly applicable across various domains, including computer vision, image processing, and surface reconstruction.

Among approximation techniques, Ordinary Least Squares (OLS) is a standard method for estimating spline coefficients but tends to be highly sensitive to outliers, leading to unreliable approximations when applied to noisy datasets. Typically, when comparing the Mean Squared Error (MSE) of a dataset contaminated by a significant number of outliers before and after their removal, a substantial decrease  is observed. This is because outliers, being far from the expected trend, contribute disproportionately to the error due to the squaring of residuals.

To address this issue, various robust spline approximation techniques have been developed, incorporating methods such as penalization terms, adaptive weighting, and statistical regularization \cite{HaTiFr09}. Among them, thin-plate splines \cite{Duchon77} provide smooth interpolation by minimizing a bending energy functional, making them effective for irregularly spaced data.  More generally, penalized regression splines introduce a regularization term to control overfitting and enhance stability \cite{Eilers96,LaWa13}. While these techniques improve robustness, they remain sensitive to extreme outliers and often require careful selection of tuning parameters.

Another common strategy for mitigating the influence of outliers is to replace the squared residuals in the cost function. For example, Least Absolute Deviations (LAD) minimizes absolute differences instead of squared differences \cite{LW04}, while the Huber loss blends squared loss for small residuals with absolute loss for large residuals \cite{Hu81, HaTiFr09}. 


In this work, we take a different approach. Rather than replacing the MSE with an alternative error measure, we introduce weights to scale each residual and optimize them using the principle of maximum entropy  in information theory and statistical mechanics  \cite{Sh48,Ja57}. This method, called Maximum Entropy Weighted Least Squares (MEWLS),  adaptively downweights anomalous data points while preserving meaningful structures in the data, enhancing robustness without requiring outlier removal. While entropy-based weighting has previously been applied to univariate settings and curve fitting \cite{GiIa21, BrGiIaRu24, DeFaIaLoMaRu25, AmFaGrIaLaMaNoRu}, this work extends the methodology to the more challenging setting of bivariate spline surfaces in multidimensional data fitting problems.

Compared to classical robust techniques such as M-estimators or Huber regression \cite{Hu81, Maronna06}, MEWLS offers several distinctive advantages. It determines the weights through an entropy maximization principle, avoiding heuristic updates or reliance on specific loss functions. This leads to a fully data-driven and parameter-free procedure that automatically adapts to the structure of the data. Moreover, MEWLS retains the interpretability and simplicity of the least-squares framework while achieving greater robustness to outliers and noise. A comparison with thin-plate splines is presented in Section 4, while additional benchmarks in the univariate setting, including comparisons with other robust techniques such as RANSAC, rlowess, rloess, Gaussian Process Regression, and Huber-based estimators, are available in \cite{BrGiIaRu24,DeFaIaLoMaRu25}.

A related framework is Iteratively Reweighted Least Squares \cite{HoWe77,OL90}, where weights are updated inside an iteration  process that progressively reduce the influence of outliers while refining the model approximation.  A recent variant  applied to multivariate adaptive spline approximation was proposed in \cite{GiImKrLoMoVi24} with the aim of preserving key data features while suppressing noise and anomalies. For further recent works exploring  robust spline-fitting strategies  see \cite{Ka23, TeBaSuCa13, TeSa13, BrGiPaSe25} and reference therein.

The paper is structured as follows. Section \ref{sec_entropy} introduces the mathematical formulation of our entropy-based bivariate spline approximation. Section \ref{sec_results} provides some theoretical results concerning the solvability of the problem and the convergence properties of the nonlinear iteration scheme employed. Section \ref{sec_example} presents numerical illustrations including one comparison, while Section \ref{sec_applications} discusses applications to two real-world image processing problems. Finally, Section \ref{sec_conclusions} concludes the paper and outlines future research directions.

\section{The principle of maximum entropy in weighted least-squares bivariate spline approximation}
\label{sec_entropy}

We consider B-spline surfaces of the form  
\begin{equation}
S(u,v) = \sum_{i = 1}^{n_1} \sum_{j=1}^{n_2} P_{ij} B_{ij}^{(d_1,d_2)} (u,v),
\label{B-spline_surf}
\end{equation}
where the coefficients $ P_{ij} \in \mathbb{R}^s $, with $ s $ typically set to $3$ for surfaces and $1$ for spline functions, represent the $ n_1 \times n_2 $ control points defining the grid on which the surface lies.\footnote{The case $s>3$, which corresponds to a 2-manifold in $\mathbb{R}^s$, is also accounted for in the subsequent computations.} In the following, it is convenient to treat elements of $\mathbb{R}^s$ as row vectors, while all other vectors will be considered, as usual, column vectors. The bivariate basis functions, defined as  
\begin{equation}
B_{ij}^{(d_1,d_2)} (u,v) = b_{i,d_1}(u) \, b_{j,d_2}(v),
\end{equation}  
form a B-spline basis for surfaces obtained through the tensor product of univariate B-spline basis functions, where $ d_1 $ and $ d_2 $ denote their respective degrees. The variables $u$ and $v$ are typically defined within the unit square $[0, 1] \times [0, 1]$.

As usual, each B-spline basis function $b_{i,d}(t)$, $i=1,\dots,n$, is defined starting from a set of ordered knots
$$ 
0 = \ts_1 \le \ts_2 \le \dots \le \ts_k = 1, 
$$
where $k=n+d+1$ determines the number of control points $n$ in relation to the degree $d$. The basis functions are constructed using the following recursion:
$$
b_{i,0} (t) =
\begin{cases}
1, & t \in [\ts_i, \ts_{i+1}), \\
0, & \text{otherwise},
\end{cases} 
$$
for $i=1,\dots,k-1$ with $b_{k-1,0} (\ts_{k})=1$, and 
$$ b_{j,d}(t) = \frac{t-\ts_j}{\ts_{j+d}-\ts_j} b_{j,d-1}(t) + \frac{\ts_{j+d+1}-t}{\ts_{j+d+1}-\ts_{j+1}} b_{j+1,d-1}(t). 
$$

Summarizing, to obtain a B-spline surface \eqref{B-spline_surf} we need to define: 
\begin{itemize}
\item an $n_1 \times n_2$ block matrix of control points in $\RR^s$ (each block $P_{ij}$ being a row vector of length $s$, according to our assumption);
\item the degrees $d_1$ and $d_2$ of the B-spline curves along the $u$ and $v$ directions;
\item the knots $\us_i$ for $i=1,\dots,n_1+d_1+1$ and $\vs_j$ for $j=1,\dots,n_2+d_2+1$ associated to the B-spline curves.
\end{itemize}
Similarly to B-spline curves, a B-spline surface is well-defined (it defines a partition of unity) in the rectangle $[\us_{d_1+1}, \us_{n_1+1}] \times [\vs_{d_2+1}, \vs_{n_2+1}]$.

In this paper, we set $d_1 = d_2 \equiv d$, and thus simplify the notation by avoiding superscripts in the basis functions. We use equispaced knots, except for the first and last ones, which are chosen to satisfy specific properties of the B-spline surface. For instance, by setting the first and last $d+1$ knots associated with each B-spline function equal to 0 and 1, respectively, the B-spline surface interpolates the corner control points.

Consider 
\begin{itemize}
\item[-] a set of $m$ input points  
\begin{equation}
(u_k, v_k, Q_k), \qquad k = 1, \dots, m,
\label{non_struct_points}
\end{equation}
with $u_k$ and $v_k$  real parameters within the domain of the B-spline surface and $Q_k\in \RR^s$;
\item[-] a sequence of positive weights $w_k$, $k=1,\dots,m$,  satisfying the normalization condition  
\begin{equation}
\sum_{k=1}^m w_k = 1.
\label{normalization}
\end{equation}
\end{itemize}
The weighted least squares problem consists in determining the control points $P_{ij}\in \RR^s$ that minimize the Mean Squared Error
\begin{equation}
\MSE = \sum_{k=1}^m w_k \| S(u_k,v_k) - Q_k \|_2^2,
\label{MSE}
\end{equation}

The least squares spline approximation is then obtained by setting to zero the partial derivatives of $\MSE$ with respect to $P_{rt}$, for $r=1,\dots,n_1$ and $t=1,\dots,n_2$. This leads to the equations  
$$
\sum_{k=1}^m w_k B_{rt}(u_k,v_k) S(u_k,v_k) = \sum_{k=1}^m w_k B_{rt}(u_k,v_k) Q_k,
$$
which are equivalent to  
$$
\sum_{i=1}^{n_1} \sum_{j=1}^{n_2} \left( \sum_{k=1}^m w_k B_{rt}(u_k,v_k) B_{ij}(u_k,v_k) \right) P_{ij} = \sum_{k=1}^m w_k B_{rt}(u_k,v_k) Q_k.
$$
Rewriting in matrix form, we obtain the system  
\begin{equation} 
{\cal B}^T \, {\cal W} \, {\cal B} \, {\cal P} = {\cal B}^T \, {\cal W} \, {\cal Q},
\label{comp_syst} 
\end{equation}
where 
\begin{itemize}
    \item ${\cal Q}$ is a $m \times s$ matrix containing the input points $Q_i$: each point is stored on a different row;
    \item ${\cal P}$ is a $n_1n_2 \times s$ matrix containing the unknown control points $P_{ij}$ reshaped column-wise and organized as the ${\cal Q}$ matrix;
    \item ${\cal W}$ is a $m \times m$ diagonal matrix containing the weights $w_i$;
    \item ${\cal B}$ is a $m \times n_1n_2$ matrix defined as
    $$
    \mathcal{B} = \left[B_{11}({\bf u},{\bf v}), B_{21}({\bf u},{\bf v}), \dots, B_{n_11}({\bf u},{\bf v}), B_{12}({\bf u},{\bf v}), \dots, B_{n_12}({\bf u},{\bf v}), \dots, B_{n_1n_2}({\bf u},{\bf v})\right]
    $$
    
 where ${\bf u}$ and ${\bf v}$ are column vectors containing $u_k$ and $v_k$ parameters, and hence each entry $B_{ij}({\bf u},{\bf v})=b_{i}({\bf u}) \cdot b_{j}({\bf v})$ is a column vector of length $m$ (the symbol $\cdot$ stands for pointwise product).  
\end{itemize}
Consequently, 
\eqref{comp_syst} is a square matrix system of size $n_1 n_2$ whose solution has $s$ columns, each of which can be computed independently using the same factorization of the coefficient matrix.
As is usual in least squares problems, \eqref{comp_syst} is equivalent to the least squares solution of the rectangular linear system
$$ {\cal W}^{1/2} \, {\cal B} \, {\cal P} = {\cal W}^{1/2}{\cal Q}. $$

In this paper, we are primarily interested in solving problems with structured input points 
\begin{equation}
(u_k, v_l, Q_{kl}), \qquad k = 1, \dots m_1 \quad \mbox{and} \quad l=1,\dots, m_2. 
\label{struct_points} 
\end{equation}
In this case, the MSE in \eqref{MSE} simplifies to  
\begin{equation}
 \MSE = \sum_{k=1}^{m_1} \sum_{l=1}^{m_2} w_{kl} \| S(u_k,v_l) - Q_{kl} \|_2^2 
\label{min_quad2} \end{equation}
with the normalization condition  \eqref{normalization} becoming
$$ \sum_{k=1}^{m_1} \sum_{l=1}^{m_2} w_{kl} = 1. $$

Repeating the same computations as before, we obtain a linear system analogous to the previous one, 
$$ \sum_{i=1}^{n_1} \sum_{j=1}^{n_2} \left( \sum_{k=1}^{m_1} \sum_{l=1}^{m_2} w_{kl} B_{rt}(u_k,v_l) B_{ij}(u_k,v_l) \right) P_{ij} = \sum_{k=1}^{m_1} \sum_{l=1}^{m_2} w_{kl} B_{rt}(u_k,v_l) Q_{kl}, $$
which can be expressed in the same matrix form as \eqref{comp_syst}, where $\cal P$ remains unchanged, and 
\begin{itemize}
    \item ${\cal Q}$ is a $m_1 m_2 \times s$ matrix defined similarly to ${\cal P}$;
    \item ${\cal W}$ is a $m_1 m_2 \times m_1 m_2$ diagonal matrix containing the weights $w_{kl}$ reshaped column-wise;
    \item $\mathcal{B}$ is the $m_1 m_2 \times n_1 n_2$ matrix 
    $$
    \mathcal{B} = [\vc(B_{11}({\bf u},{\bf v})), \vc(B_{21}({\bf u},{\bf v})), \dots, \vc(B_{n_11}({\bf u},{\bf v})), \vc(B_{12}({\bf u},{\bf v})), \dots]
    $$
    where now $\mathbf{u}$ and $\mathbf{v}$ are column vectors of length $m_1$ and $m_2$, respectively, and  $B_{ij}(\mathbf{u}, \mathbf{v}) = b_i({\bf u}) b_{j}({\bf v}^T)$ is a matrix of size $m_1 \times m_2$, while $\vc$ is the matrix vectorization operation that reshapes $B_{ij}(\mathbf{u}, \mathbf{v})$ in a column-wise order. 
\end{itemize}

We now address the problem of determining the weights to account for the presence of anomalous data.\footnote{In the following, we focus on the general problem \eqref{MSE}. The adaptation to problem \eqref{min_quad2} is straightforward.}  
Given the normalization condition \eqref{normalization}, the weight distribution \(\{w_i\}\) can be interpreted as a probability distribution, allowing us to define the associated entropy as  

\begin{equation}  
\label{entropy}  
H(w_1,\dots,w_m) = -\sum_{i=1}^m w_i \log w_i.  
\end{equation}

In information theory, \eqref{entropy} quantifies the expected information content (or uncertainty) of the distribution \(\{w_i\}\). To determine the weights, we adopt the well-known principle of maximum entropy,  which states that, among all distributions satisfying a given set of constraints, the one with highest entropy should be selected. In fact, this choice  ensures that, apart from the imposed constraints, no additional assumptions are introduced, thus preventing any bias that might arise from incorporating information that is not actually available.  

To determine the appropriate constraints in our context, we first note that maximizing \eqref{entropy} subject only to \eqref{normalization} yields the uniform distribution $ w_i = 1/m $, which corresponds to Laplace's principle of indifference. This principle states that, in the absence of prior knowledge favoring any particular outcome, probabilities should be assigned equally among all possibilities. In our setting, this  leads to the Ordinary Least Squares (OLS) approximation, and we denote by $\MSEuw$ the resulting Main Squared Error  ($\mathrm{uw}$ stands for \textit{uniform weights}).  

As noted in the introduction, OLS is highly sensitive to outliers: their large residuals may significantly increase the MSE. If we had a way to  assign these outliers a negligible weight, we would expect a decrease in the error. This observation motivates the introduction of an additional constraint: we enforce the weighted Mean Squared Error to take a prescribed value $\MSE \leq \MSEuw$.  The entropy maximization process still tries to spread the weights as evenly as possible, but now it must also respect the constraint that the weighted error sum stays at $\MSE$.  As we will see below, the natural outcome of this approach is that outliers receive very small weights, which causes an effective reduction of their influence on the overall approximation.  Summarizing, the maximal entropy approach requires the solution of the following constrained optimization problem: 
\begin{equation}
\label{constr}
\begin{array}{rcl}
\mbox{maximize}     &&\displaystyle   -\sum_{i=1}^m w_i \log w_i, \\[.5cm] 
\mbox{subject to:}  && \displaystyle \sum_{i=1}^m w_i = 1,   \\[.5cm] 
                    && \displaystyle \sum_{i=1}^m w_i \| S(u_i,v_i) - Q_i \|_2^2 =  \MSE,   
\end{array}
\end{equation}
where $S(u,v)$ is defined at (\ref{B-spline_surf}) and $\MSE$ takes a prescribed input value. 

To solve this constrained optimization problem  we employ the method of Lagrange multipliers. We introduce the multipliers $\lambda $ and $\mu$ corresponding to the constraints and define the Lagrangian function as  

\begin{equation}
\label{Lagrangian}
\begin{array}{rl}
\mathcal{L}(w_1, \dots, w_m, P_{11}, \dots, P_{n_1n_2}, \lambda, \mu) = & \displaystyle
- \sum_{i=1}^{m} w_i \log w_i 
+ \lambda \left( 1 - \sum_{i=1}^{m} w_i \right) \\[.4cm]
&\displaystyle + \mu \left( \MSE - \sum_{i=1}^{m} w_i \left\| \sum_{r = 1}^{n_1} \sum_{s=1}^{n_2} P_{rs} B_{rs}(u_i,v_i) - Q_i \right\|_2^2 \right).
\end{array}
\end{equation}
To find the stationary points and impose the first-order optimality conditions, we compute the partial derivatives of $\mathcal{L}$ with respect to the arguments and set them to zero. 
Differentiating with respect to $w_k$ yields
\begin{equation}
\label{Dwk}
\frac{\partial \mathcal{L}}{\partial w_k} = - (1 + \log w_k) - \lambda - \mu \left\| \sum_{r=1}^{n_1} \sum_{s=1}^{n_2} P_{rs} B_{rs}(u_k,v_k) - Q_k \right\|_2^2=0.
\end{equation}
For computational purposes, we rearrange (\ref{Dwk}) to express $w_k$ explicitly:
\begin{equation}
\label{wk1}
w_k = \exp(-1 - \lambda) \, \exp\left( - \mu \left\| \sum_{r = 1}^{n_1} \sum_{s=1}^{n_2} P_{rs} B_{rs}(u_k,v_k) - Q_k \right\|_2^2\right).
\end{equation}
Next, differentiating with respect to $P_{ij}$ gives
\begin{equation}
\frac{\partial \mathcal{L}}{\partial P_{ij}} = -2 \mu \sum_{k=1}^{m} w_k B_{ij}(u_k,v_k) \, \left( \sum_{r=1}^{n_1} \sum_{s=1}^{n_2} P_{rs} B_{rs}(u_k,v_k) - Q_k \right) = 0.
\end{equation}
This corresponds to the normal system (\ref{comp_syst}). Finally, setting the derivatives with respect to $\lambda$ and $\mu$ to zero  yields the two constraints in (\ref{constr}):
\begin{equation}
\label{constr1}
\sum_{i=1}^{m} w_i - 1= 0
\end{equation}
and
\begin{equation}
\label{constr2}
\sum_{i=1}^{m} w_i \left\| \sum_{r = 1}^{n_1} \sum_{s=1}^{n_2} P_{rs} B_{rs}(u_i,v_i) - Q_i \right\|_2^2 -\MSE = 0.
\end{equation}
We can remove the unknown $\lambda$ summing over equations (\ref{wk1}) and using (\ref{constr1}). We obtain
$$
\exp(1 + \lambda)= \sum_{i=1}^m\exp\left( - \mu \left\| \sum_{r = 1}^{n_1} \sum_{s=1}^{n_2} P_{rs} B_{rs}(u_i,v_i) - Q_i \right\|_2^2\right) = \sum_{i=1}^m \exp\left( - \mu \left\| S(u_i,v_i) - Q_i \right\|_2^2\right) .
$$
Substituting this expression  in (\ref{wk1}) and inserting (\ref{wk1}) into  (\ref{constr2}), the final system of equations we need to solve takes the form
\begin{subequations}
\label{sys}
\begin{flalign}
{\cal B}^T{\cal W}{\cal B} \, {\cal P} = {\cal B}^T{\cal W}{\cal Q}, \label{sysa} \\[.3cm]
\sum_{i=1}^{m}  \exp\left( - \mu \left\| S(u_i,v_i) - Q_i \right\|_2^2\right) \left\| S(u_i,v_i) - Q_i \right\|_2^2 
- \MSE \cdot \sum_{i=1}^m\exp\left( - \mu \left\| S(u_i,v_i) - Q_i \right\|_2^2\right) = 0, \label{sysb} \\[.3cm]
\displaystyle w_k = \frac{\displaystyle \exp\left( - \mu \left\| S(u_k,v_k) - Q_k \right\|_2^2\right)}{ \displaystyle \sum_{i=1}^m\exp\left( - \mu \left\| S(u_i,v_i) - Q_i \right\|_2^2\right)}, \qquad k=1,\dots,m. \label{sysc}
\end{flalign}
\end{subequations}
We may interpret system (\ref{sys}) as consisting of three subsystems:
\begin{itemize}
\item[-] The normal system (\ref{sysa}) determines the unknowns $P_{ij}$. This system is linear with respect to the unknown vector $\cal P$.
\item[-] The nonlinear scalar equation (\ref{sysb}) determines the unknown $\mu$. 
\item[-] The final set of $m$ explicit equations in (\ref{sysc}) provides  the vector of weights $w=(w_1,w_2,\dots,w_m)^\top$. 
\end{itemize}
The derived weight formula \eqref{sysc} shows that outliers, i.e., points where the residual $\left\| S(u_k,v_k) - Q_k \right\|_2^2$ is large, receive exponentially smaller weights. Meanwhile, the update equation for $P_{ij}$ (\ref{sysa}) ensures that the spline representation is adjusted in accordance with the weighted data distribution.
Clearly, the three subsystems (\ref{sysa})–(\ref{sysc}) are not independent, as each involves the full set of unknowns. 
It should be emphasized that, due  to its nonlinear nature, system \eqref{sys} may generally admit  multiple solutions. 
Since our ultimate goal is to improve the accuracy of the smoothing spline obtained with uniform weights, we employ a continuation technique: the value of $\MSE$ in \eqref{constr} is initially set to $\MSEuw$ and then gradually reduced until it reaches the desired terminal value.

\section{Implementation aspects and  theoretical results}
\label{sec_results}
The above interpretation of (\ref{sys}) suggests that a natural way to solve it numerically is through a nonlinear Gauss--Seidel method that updates the approximation  $({\cal P}^{(j)},\mu^{(j)},{w}^{(j)})$ at step $j$  by first solving the linear system  (\ref{sysa}) to define ${\cal P}^{(j+1)}$. This is subsequently inserted in the scalar equation (\ref{sysb}) to produce $\mu^{(j+1)}$ and, finally, ${\cal P}^{(j+1)}$ and $\mu^{(j+1)}$ are plugged in (\ref{sysc}), from which the new weight distribution ${w}^{(j+1)}$ is retrieved. A clear advantage is that, despite each block equation involving nonlinear terms, the only nonlinearity inherited by the Gauss--Seidel  process amounts to a single scalar equation, namely (\ref{sysb}), which may be easily handled by a Newton-like method. The whole iteration procedure is  summarized in Algorithm \ref{Alg1}. 
\begin{algorithm}
\begin{algorithmic}[1]
\State for $k=1,\dots,m$, set  $w_k \gets\frac{1}{m}$ (uniform weights), $\mu_0 \gets 0$
\Repeat:
\State $\displaystyle  \cal P \gets$ solution of the least squares problem ${\cal W}^{1/2} {\cal B} \, {\cal P} = {\cal W}^{1/2}{\cal Q}$ 
\State 
$\mu \gets$ solution of equation (\ref{sysb}) via a Newton-like iteration scheme with initial guesses $\mu_0$  \\
 \hspace*{1.2cm} and ${\cal P}$  computed at  step 3 \vspace*{.2cm}
\State 
$\mu_0 \gets \mu$
\State $\displaystyle w_k \gets {\displaystyle \exp\left( - \mu \left\| S(u_k,v_k) - Q_k \right\|_2^2\right)} /{ \displaystyle \sum_{i=1}^m\exp\left( - \mu \left\| S(u_i,v_i) - Q_i \right\|_2^2\right)}$, $k=1,\dots,m$ \vspace*{.2cm}
\Until convergence to within a prescribed tolerance is attained
\end{algorithmic}
\caption{Numerical procedure for  solving system (\ref{sys})}
\label{Alg1}
\end{algorithm}

In this section, we discuss the solvability of system (\ref{sys}) for values of the parameter $\MSE$ lying in a left neighborhood of $\MSEuw$. Furthermore, we establish the convergence of the iterative scheme described in Algorithm~\ref{Alg1}. Although these results are local in nature, they represent a preliminary step toward a more comprehensive analysis, which will require more sophisticated tools and will be addressed in a broader setting in future work. 

For the sake of simplicity, and without loss of generality, we consider the case where (\ref{B-spline_surf}) represents a spline function, so that $s = 1$, ${\cal Q} \in \mathbb{R}^m$, and ${\cal P} \in \mathbb{R}^{n}$, with $n=n_1n_2$.  For a generic real-valued function $g: \mathbb{R} \rightarrow \mathbb{R}$ and a vector $z \in \mathbb{R}^q$, we define $g(z) := (g(z_1), \dots, g(z_q))^\top$. Hence, for example, $z^2 = z \odot z = (z_1^2, \dots, z_q^2)^\top$, where $\odot$ is the Hadamard (elementwise) product. Furthermore, $\diag(z)$ will denote the diagonal $q\times q$ matrix whose principal diagonal is $z$.

Let $r = {\cal B}{\cal P} - Q$ be the residual vector,   $w = (w_1,\dots,w_m)^\top$ the weight vector (the principal diagonal of ${\cal W}$), $u=(1,\dots,1)^\top \in \RR^m$  the unitary vector and $I$ the identity matrix of dimension $m$. We rewrite system (\ref{sys})  as 
\begin{equation}
\label{sys1}
F({\cal P},\mu,w) = (F_1, F_2, F_3)^\top = 0,
\end{equation} 
where
\begin{subequations}
\label{sys11}
\begin{align}
F_1({\cal P}, w) &= {\cal B}^\top {\cal W} {\cal B}  {\cal P} - {\cal B}^\top {\cal W} {\cal Q}, \\
F_2({\cal P}, \mu) &= \sum_{i=1}^{m} \exp(-\mu r_i^2) r_i^2 - \MSE \sum_{i=1}^{m} \exp(-\mu r_i^2), \\
F_3({\cal P}, \mu, w) &= w - \frac{\exp(-\mu({\cal B}{\cal P} - Q)^2)}{u^\top \exp(-\mu({\cal B}{\cal P} - Q)^2)}.
\end{align}
\end{subequations}
Corresponding to the choice $\MSE =\MSEuw$ is the ordinary least squares approximation obtained with the uniform-weight distribution. In such a case, the solution of (\ref{sys1}) reads
\begin{equation}
\label{uwsolution}
({\cal P}^*, \mu^*,w^*)^\top=\left({\cal (B^\top B)}^{-1}{\cal B^\top Q}, 0,\frac{1}{m}u\right)^\top.
\end{equation}

\begin{theorem}
\label{maintheo}
Assume that the entries of the residual vector $r^* = \mathcal{B} \mathcal{P}^* - \mathcal{Q}$ do not all have the same modulus. Then, there exist $\delta > 0$ and unique functions $\mathcal{P}(\MSE)$, $\mu(\MSE)$, and $w(\MSE)$, defined on the interval $(\MSEuw - \delta, \MSEuw)$, such that:
\begin{itemize}
    \item[(a)] $F\left(\mathcal{P}(\MSE), \mu(\MSE), w(\MSE)\right) = 0$;
    \item[(b)] the nonlinear Gauss--Seidel iteration is locally convergent.
\end{itemize}
\end{theorem}
\begin{proof}
The Jacobian of $F({\cal P},\mu,w)$ takes the following form:
\begin{equation}
\label{jacF}
J_F ({\cal P},\mu,w) = 
\left(
\begin{array}{lll}
A & 0 & C \\
d^\top & s & 0^\top \\
D & v & I
\end{array}
\right),
\end{equation}
where
\begin{itemize}
\item $\displaystyle A = \frac{\partial F_1}{\partial {\cal P}} = {\cal B}^\top {\cal W} {\cal B} \in \RR^{n\times n}$,
\item $\displaystyle C = \frac{\partial F_1}{\partial {w}} = {\cal B}^\top \diag(r) \in \RR^{n\times m}$,
\item $\displaystyle d^\top = \frac{\partial F_2}{\partial {\cal P}} = \left(2r\odot((1+\mu \MSE)u -\mu r^2)\odot e^{-\mu r^2} \right)^\top {\cal B} \in \RR^{1\times n}$,
\item $\displaystyle s = \frac{\partial F_2}{\partial \mu} = \sum_{i=1}^m r_i^2 e^{-\mu r_i^2} (\MSE - r_i^2) \in \RR$,
\item $\displaystyle D = \frac{\partial F_3}{\partial {\cal P}} = \frac{2\mu}{\sum_{i=1}^m e^{-\mu r_i^2}}\diag(e^{-\mu r^2})\left( \frac{1}{\sum_{i=1}^m e^{-\mu r_i^2}}\diag(r) - u (r \odot e^{-\mu r^2})^\top  \right){\cal B} \in \RR^{m \times n}$,
\item $\displaystyle v = \frac{\partial F_3}{\partial {\mu}}  = \frac{1}{(\sum_{i=1}^m e^{-\mu r_i^2})^2} \diag(e^{-\mu r^2}) \left( \left(\sum_{i=1}^m e^{-\mu r_i^2}\right)r^2 - \left(\sum_{i=1}^m r_i^2e^{-\mu r_i^2}\right)u \right)\in \RR^{m \times 1}$,
\item $\displaystyle I = \frac{\partial F_3}{\partial {w}}$ (identity matrix).
\end{itemize}
Let us add asterisks to the above quantities when they are evaluated at the critical point (\ref{uwsolution}).

We begin by analyzing the scalar quantity
\begin{equation}
\label{s*}
s^* = \sum_{i=1}^m (r_i^*)^2 \left( \MSEuw - (r_i^*)^2 \right),
\end{equation}
where $\MSEuw=1/m \sum_{i=1}^m (r_i^*)^2$.  Substituting this expression yields
$$
s^* = \sum_{i=1}^m (r_i^*)^2 \left( \frac{1}{m} \sum_{j=1}^m (r_j^*)^2 - (r_i^*)^2 \right)
= \frac{1}{m} \sum_{i=1}^m \sum_{j=1}^m (r_i^*)^2 (r_j^*)^2 - \sum_{i=1}^m (r_i^*)^4 = \frac{1}{m} \left( \sum_{i=1}^m (r_i^*)^2 \right)^2 - \sum_{i=1}^m (r_i^*)^4.
$$
We now apply Jensen's inequality to the convex function $x \mapsto x^2$, obtaining the bound:
$$
\left( \frac{1}{m} \sum_{i=1}^m (r_i^*)^2 \right)^2 \leq \frac{1}{m} \sum_{i=1}^m (r_i^*)^4,
$$
with equality if and only if all $(r_i^*)^2$ are equal. Therefore, under the assumption that 
$|r^*|$  is not constant, the inequality is strict, and we conclude that $s^* < 0$. 

It follows that the diagonal block 
$$
\left(\begin{array}{ll}
s^* & 0^\top \\
v^* & I
\end{array}\right),
$$
is nonsingular, so the invertibility of matrix $J_F({\cal P}^*,\mu^*,w^*)$ is tantamount to the invertibility of the Schur complement
$$
S^*:=A^*-
\begin{pmatrix}
0 & C^* 
\end{pmatrix} 
\left(
\begin{array}{ll}
s^* & 0^\top \\
v^* & I
\end{array}\right)^{-1} 
\begin{pmatrix}
(d^*)^\top \\
D^*
\end{pmatrix}.
$$
Considering that
\begin{equation}
\label{starseq}
\begin{array}{l}
\displaystyle A^*=\frac{1}{m}{\cal B}^\top {\cal B}, \quad C^*={\cal B}^\top \diag(r^*), \quad v^*=\frac{1}{m}\left(m (r^*)^2-(u^\top (r^*)^2)u \right),\\[.3cm]
\displaystyle (d^*)^\top=2(r^*)^\top B=0^\top, \quad  D^*=0,
\end{array}
\end{equation}
we conclude that $S^*=A^*$ is positive definite and hence nonsingular.  Since $F$ is continuously differentiable in a neighborhood of $({\cal P}^*,\mu^*,w^*)$, by the implicit function theorem, it follows that there exists a neighborhood of $\MSEuw$ in which the solution $(\mathcal{P}, \mu, w)$ of (\ref{sys1}) exists, is unique, and  depends smoothly on $\MSE$.

Now we prove property (b). Local convergence properties of the nonlinear process described in Algorithm \ref{Alg1}  are governed by the Gauss--Siedel iteration applied to the linearization of system (\ref{sys}) in a neighborhood of the solution $({\cal P}^{\ast},\mu^{\ast},{w}^{\ast})$, namely the matrix $J_F$ defined at (\ref{jacF}) (see, for example, \cite[page 65]{Rhe98}).

Apart from the upper-right block $C$, the Jacobian $J_F$ is a block triangular matrix with two definite positive diagonal blocks: $A={\cal B}^\top {\cal W} {\cal B}$ and $I$. This structure makes the Gauss--Seidel process particularly well suited to the problem at hand. The Gauss-Seidel iteration matrix 
$$
G = - \left(
\begin{array}{lll}
A & 0 & 0 \\
d^\top & s & 0^\top \\
D & v & I
\end{array} \right)^{-1}
\left( \begin{array}{lll}
0 & 0 & C \\
0^\top & 0 & 0^\top \\
0 & 0 & 0
\end{array}\right)
$$
has nonzero entries only in the last block column. Therefore, the convergence properties of the iteration  depend entirely on the  bottom-right block $G_{33}$ which takes the form 
\begin{equation}
\label{G33}
G_{33}  = (D  + s^{-1} v  d^\top )A^{-1} C.
\end{equation}
Convergence is guaranteed if the spectral radius of this matrix is less than one. This is indeed the case when $\MSE \in (\MSEuw-\delta, \MSEuw)$ for some  sufficiently small $\delta$, since (see (\ref{starseq}))
$$
\lim_{\MSE \rightarrow \MSEuw} ||D(\MSE)|| = D^\ast = 0, \qquad  \lim_{\MSE \rightarrow \MSEuw} ||d^\top(\MSE)|| = (d^\ast)^\top = 0^\top,
$$
while $v$, $A^{-1}$,  $C$ remain bounded and $s$ is bounded away from $0$. As a result, $||G_{33}||$ becomes arbitrarily small as $\MSE$ approaches $\MSEuw$ and the iteration is convergent.
\end{proof}

For a large set of test problems, we observed that the entries of $G_{33}$	remain small even for values of
$\MSE$ significantly different from $\MSEuw$. This circumstance, combined with the use of a continuation technique on the parameter $\MSE$ that supplies a good initial guess within the basin of attraction of the scheme, has ensured that no critical convergence issues have been encountered in practice. An illustration is provided in Section \ref{sec_franke}.

\section{Illustrative examples}
\label{sec_example}
In this section, we show the effectiveness of the entropy-based spline approach using two synthetic datasets with a known ground truth, allowing for a direct assessment of accuracy and robustness. The objective is to reconstruct surfaces from noisy and outlier-contaminated data while preserving their structure. The first example examines surface approximation from a noisy point cloud, to evaluate the ability of the method to handle scattered data. The second example focuses on reconstructing a structured surface while reducing the influence of outliers.  

All numerical experiments were performed in Matlab\textsuperscript{\textregistered} (version R2024b) on a computer equipped with a 3.6 GHz Intel i9 processor and 32 GB of RAM.

\subsection{Surface approximation from a noisy point cloud}
\label{sec_franke}
We consider a point cloud obtained by randomly sampling the Franke function \cite{Fr79,SuBi13} 
$$
\begin{array}{rcl}
F(x,y) &=& \displaystyle \frac34 e^{-\left( (9x-2)^2+(9y-2)^2 \right)/4} + \frac34 e^{-\left( (9x+1)^2/49+(9y+1)/10 \right)} \\[.3cm]
       & & \displaystyle + \frac12 e^{-\left( (9x-7)^2+(9y-3)^2 \right)/4} - \frac15 e^{-\left( (9x-4)^2+(9y-7)^2 \right)}
\end{array}
$$
over its domain $[0, 1] \times [0, 1]$. The function is positive, has two relative maxima and one minimum, and takes values in $[0,1]$ except for a small region surrounding one of the peaks.  More specifically, the input dataset consists of a total of 1150 points:  
\begin{itemize}
\item[(i)] $1000$  points  perturbed by a Gaussian noise with a  standard deviation $\sigma=10^{-3}$ relative to the function values; 
\item[(ii)] $150$ points randomly distributed within the cube $[0, 1] \times [0, 1] \times [0, 1]$. Excluding points that happen to be near the Franke surface, we expect that almost all of these points behave as outliers.      
\end{itemize}
To approximate the input data, we consider a cubic B-spline surface defined on a $10 \times 10$ grid. The knots are selected to ensure interpolation at the four corners of the grid, regardless of whether the dataset contains points at those positions. The variables $u$ and $v$ associated with the point cloud are set equal to their respective $x$- and $y$-coordinates. As a result, it is sufficient to compute only the $z$-values of the control points, meaning that $s=1$.  

\begin{figure}
\includegraphics[width=.5\textwidth]{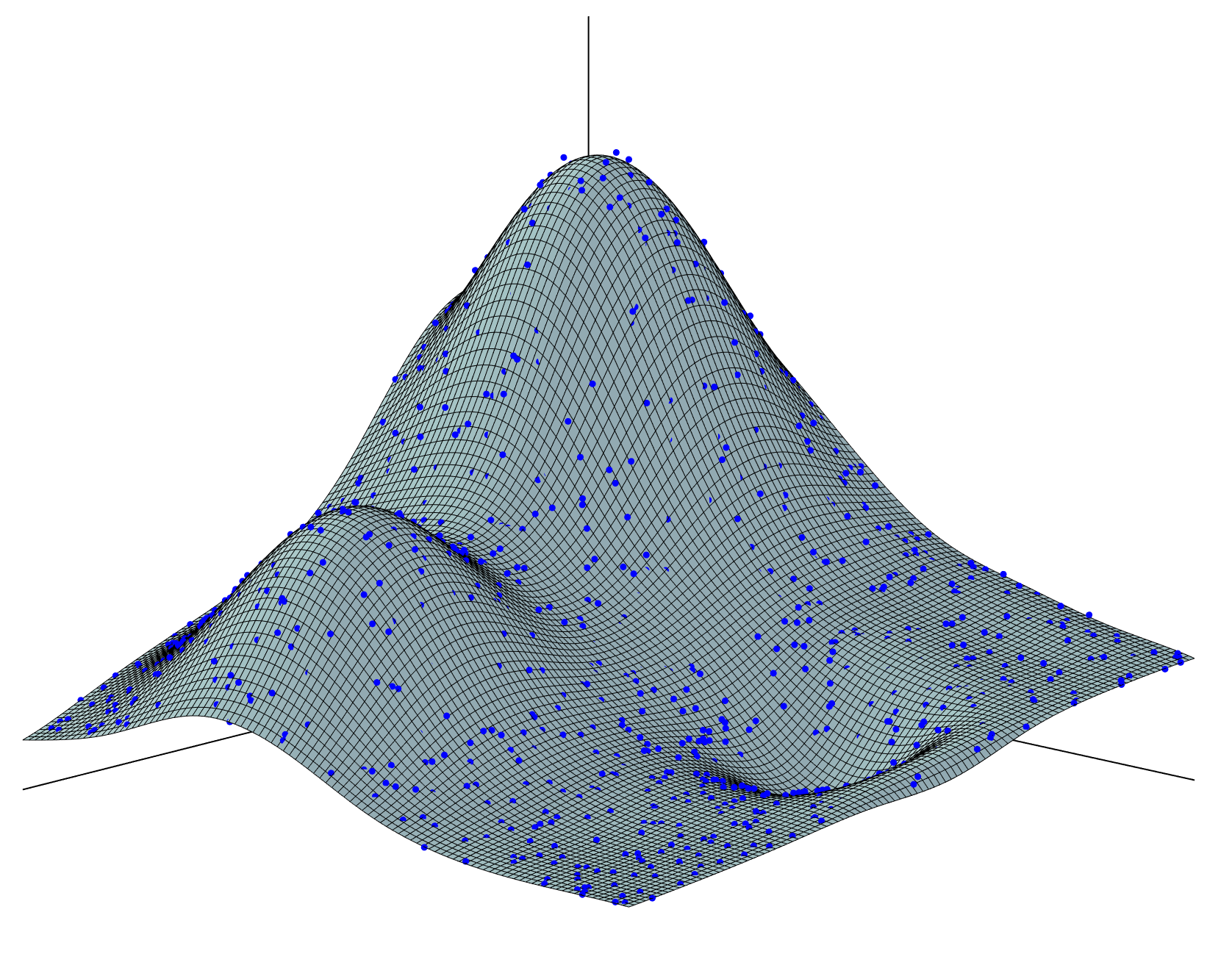} 
\includegraphics[width=.5\textwidth]{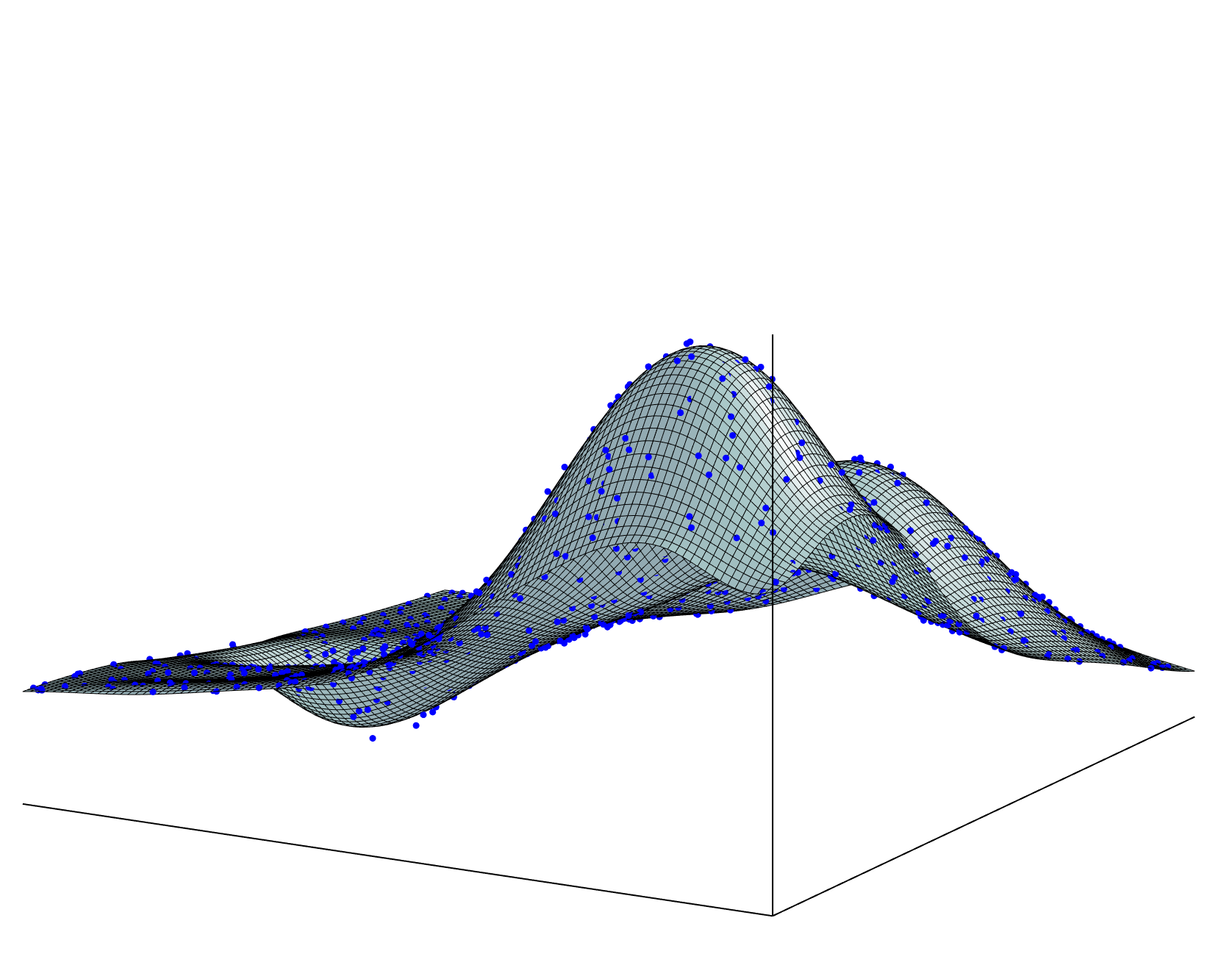} 
\caption{Two views of the least squares approximation of the 1000 randomly sampled points from case (i) (outliers from case (ii) are excluded).  \label{fig_franke1}}
\end{figure}

\begin{figure}
\includegraphics[width=.5\textwidth]{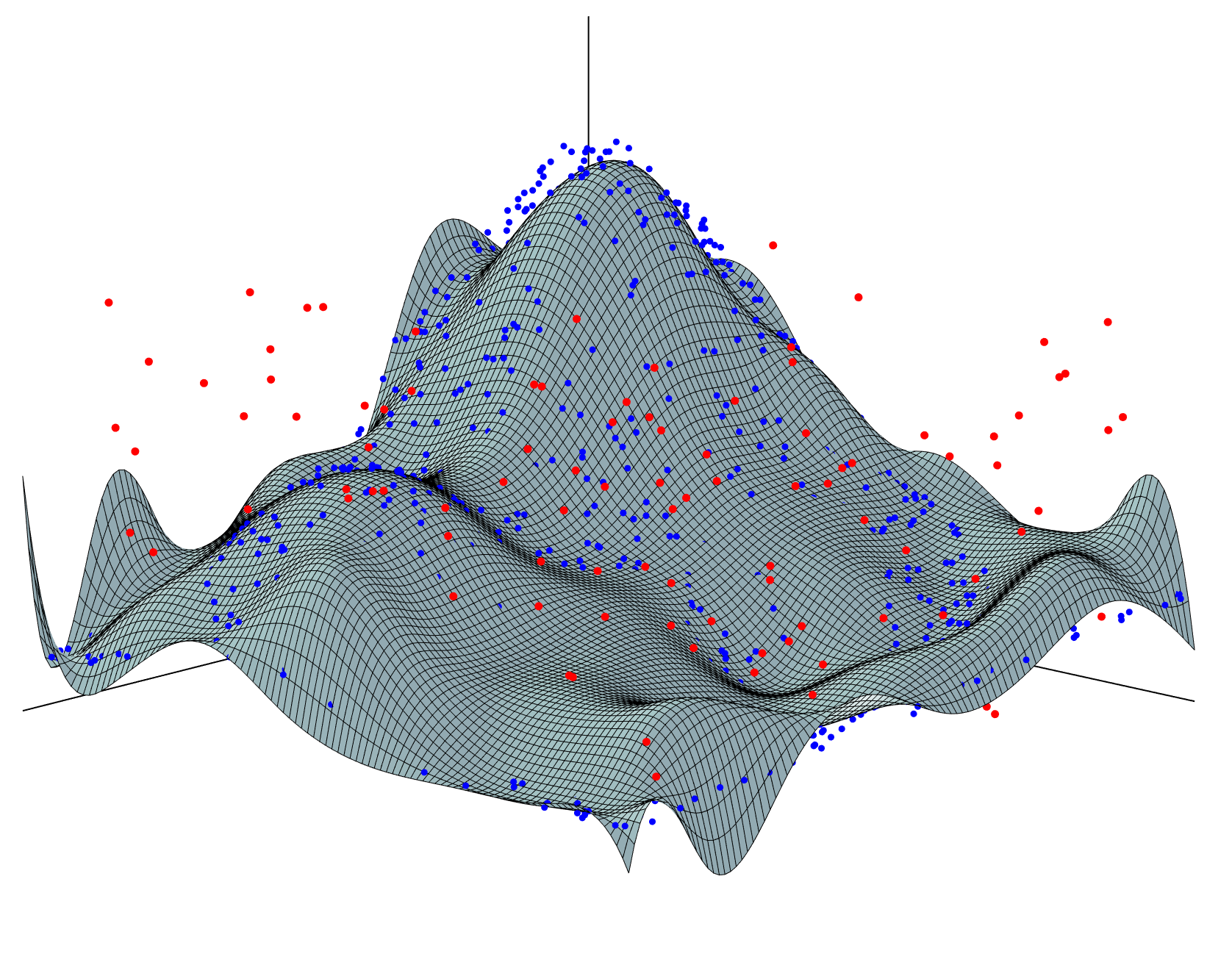} 
\includegraphics[width=.5\textwidth]{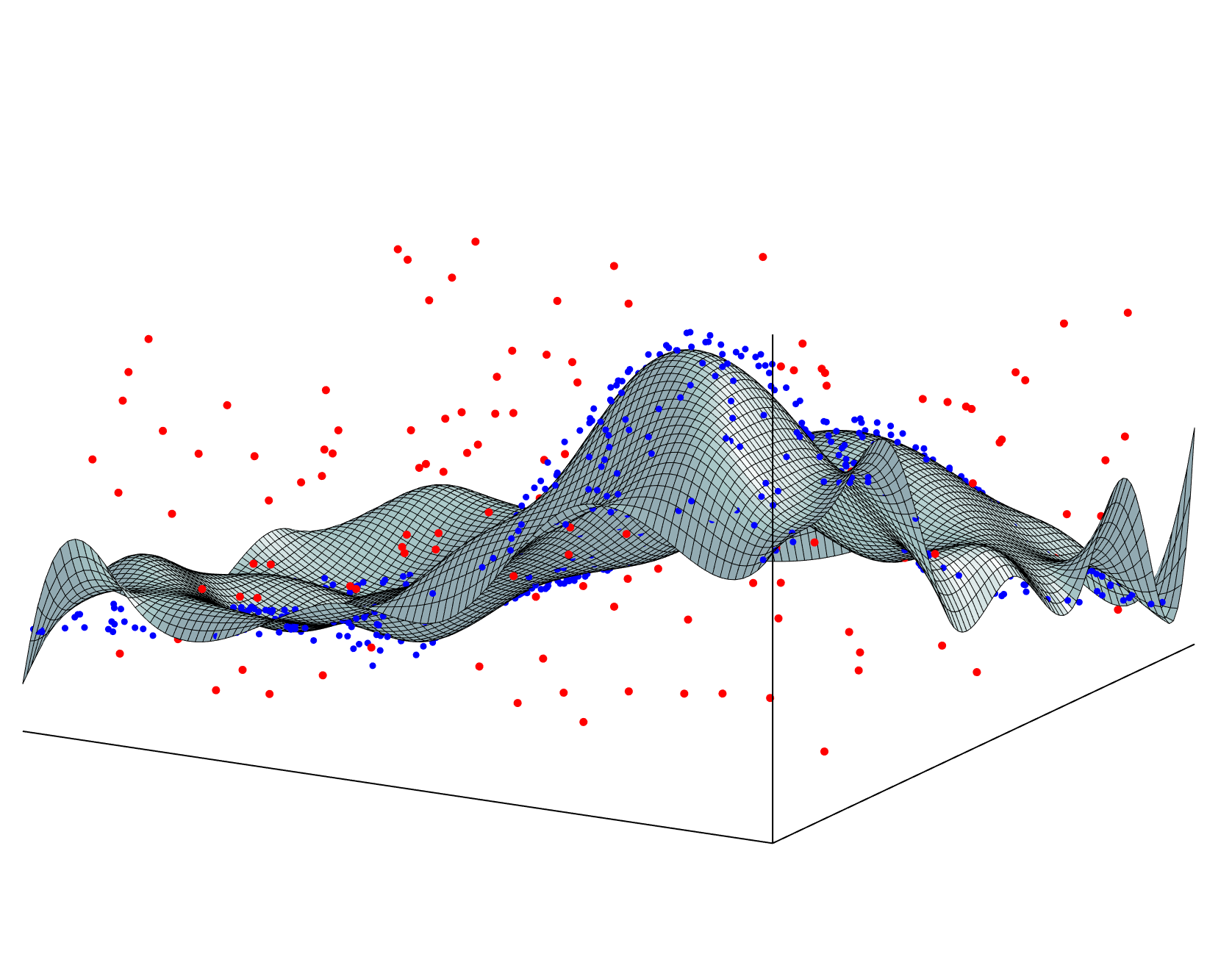} 
\caption{Least squares approximation from the same view angles as in Figure \ref{fig_franke1}, considering the entire dataset of all $1150$ points. Outliers are shown in red. \label{fig_franke2}}  
\end{figure}

To assess the impact of outliers, we begin by examining the OLS approximation under two different scenarios: 
\begin{itemize}
\item[(a)] In the absence of outliers (i.e., points from (ii) are excluded), the OLS approximation yields a mean squared error $\MSEuw = 1.83 \cdot 10^{-3}$. To assess its accuracy in reproducing the Franke function through cross-validation, we compute the mean squared error of the spline surface with respect to $F(x,y)$, evaluated on a uniform grid $\Omega$ of size $101\times 101$ covering the domain $[0,1]\times[0,1]$, obtaining $\MSEcv = 2.37 \cdot 10^{-5}$. This confirms that the resulting surface provides an accurate approximation of the Franke function (see Figure \ref{fig_franke1}).  

\item[(b)] The presence of outliers drastically alters the least squares approximation yielding a mean squared errors $\MSEuw=1.76 \cdot 10^{-2}$ and  $\MSEcv=3.61 \cdot 10^{-3}$, this latter being two orders of magnitude greater than its corresponding value in case (a) (see Figure \ref{fig_franke2}). In additional experiments (not reported here), significantly larger peaks appear at the corners of the domain, which are not present in the original function.  
\end{itemize}

\begin{figure}
\includegraphics[width=.5\textwidth]{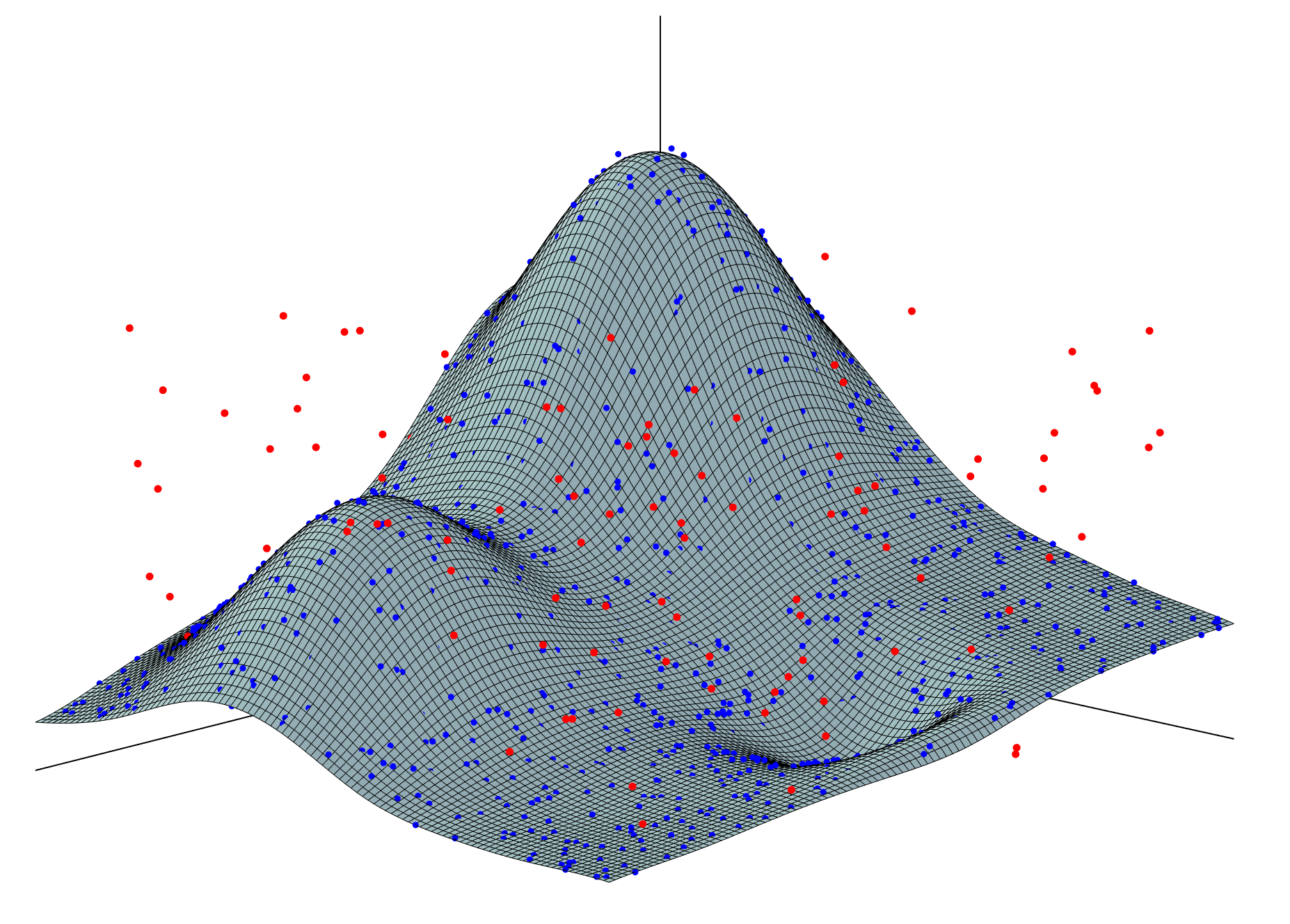} 
\includegraphics[width=.5\textwidth]{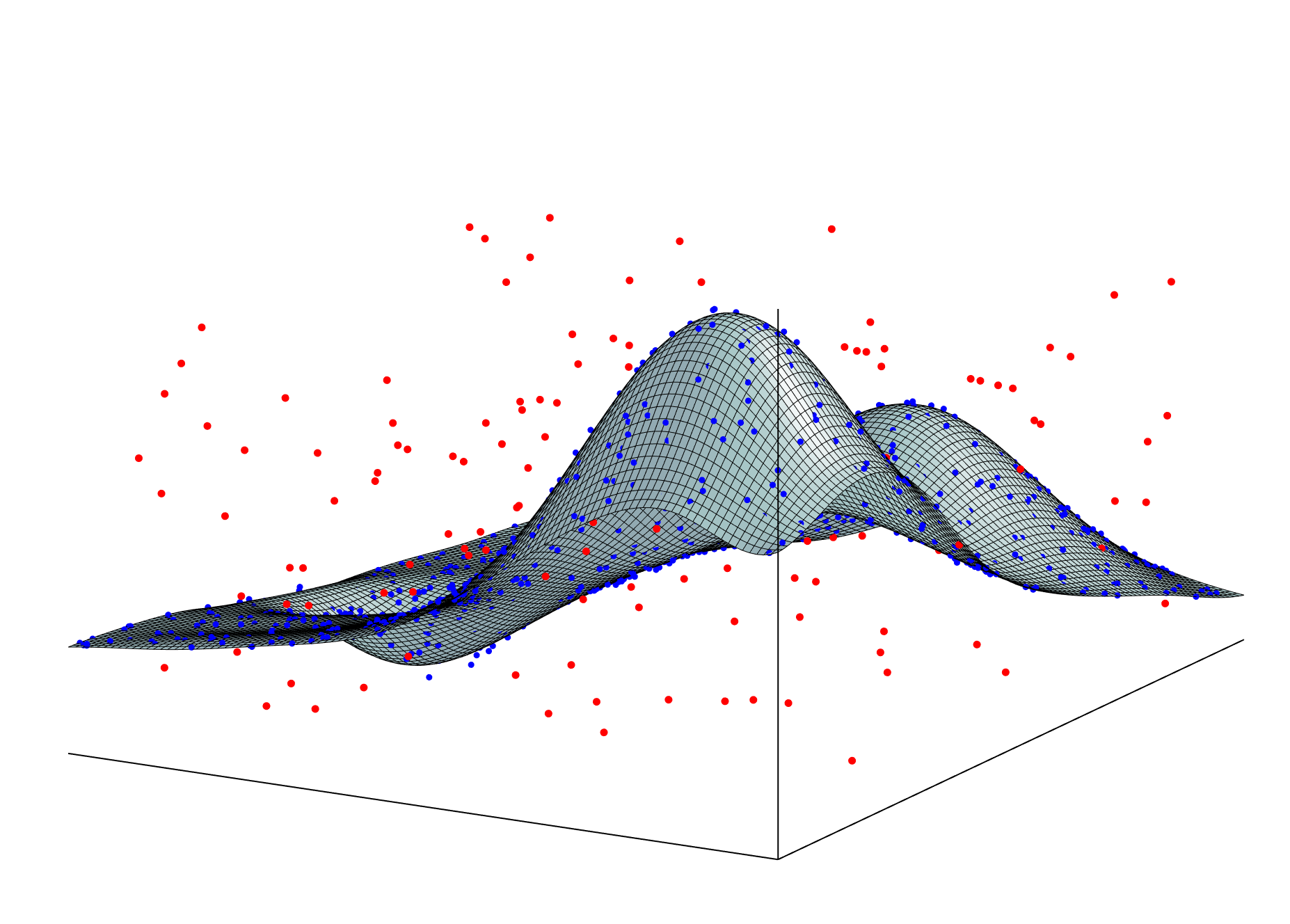} 
\caption{Maximum entropy least squares approximation ($r=500$) considering the entire dataset. Outliers are shown in red. Compare with Figure \ref{fig_franke1}. \label{fig_franke3}}  
\end{figure}

Concerning the MEWLS approximation, we solve \eqref{constr} by selecting two different values of $\MSE$:  
\begin{equation}
\MSE = \MSEuw / r, \quad r = 2, 500,
\label{scal_fac}
\end{equation}
where $\MSEuw$ corresponds to the mean squared error for the complete dataset (case (b) above). Figure \ref{fig_franke3} shows the approximation surface for $r = 500$, which closely resembles the one in Figure \ref{fig_franke1}. Indeed, the mean squared error with respect to the Franke function evaluated at $\Omega$ is now $\MSEcv = 2.55 \cdot 10^{-5}$, consistent with case (a) above.  

\begin{figure}
\includegraphics[width=.5\textwidth]{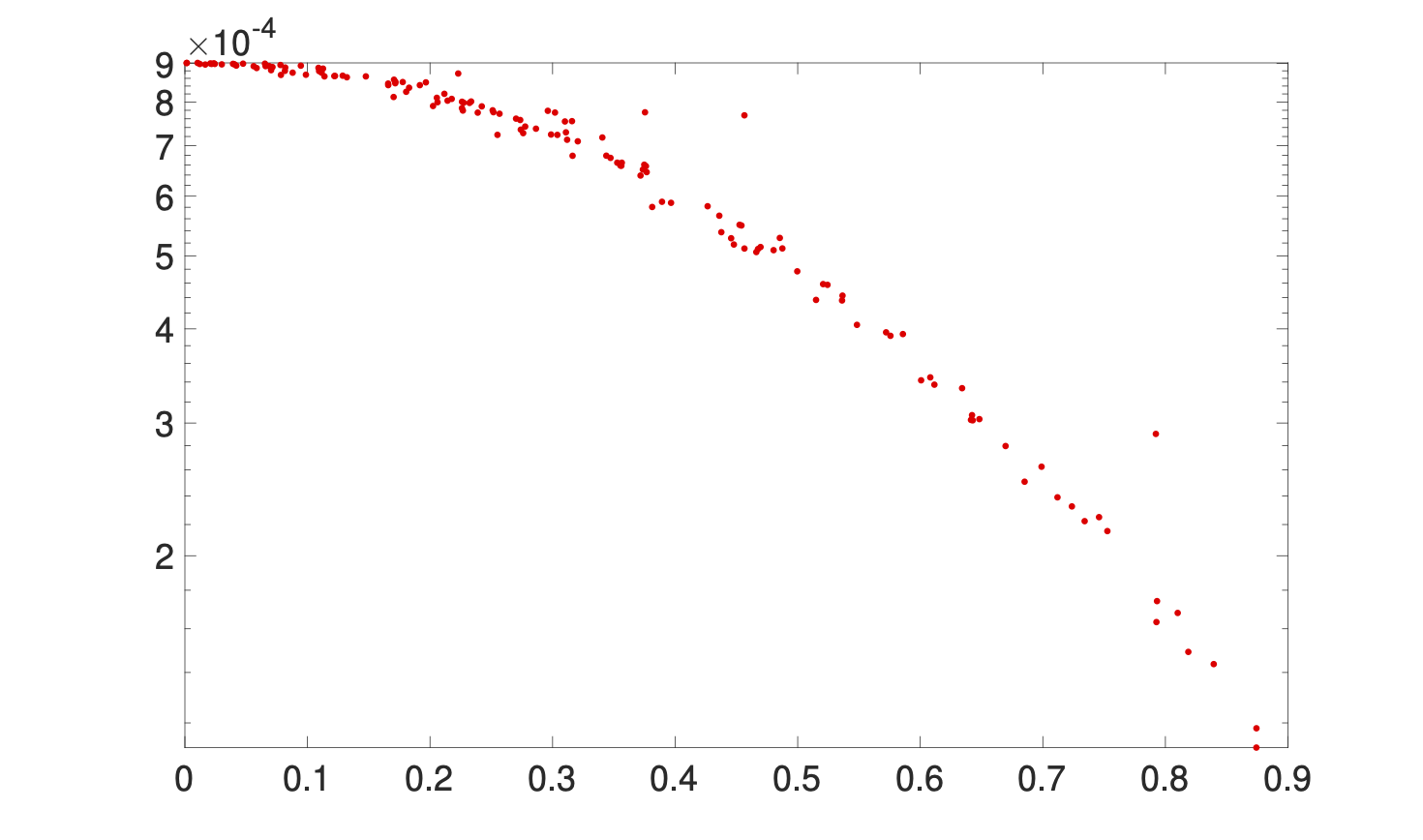} 
\includegraphics[width=.5\textwidth]{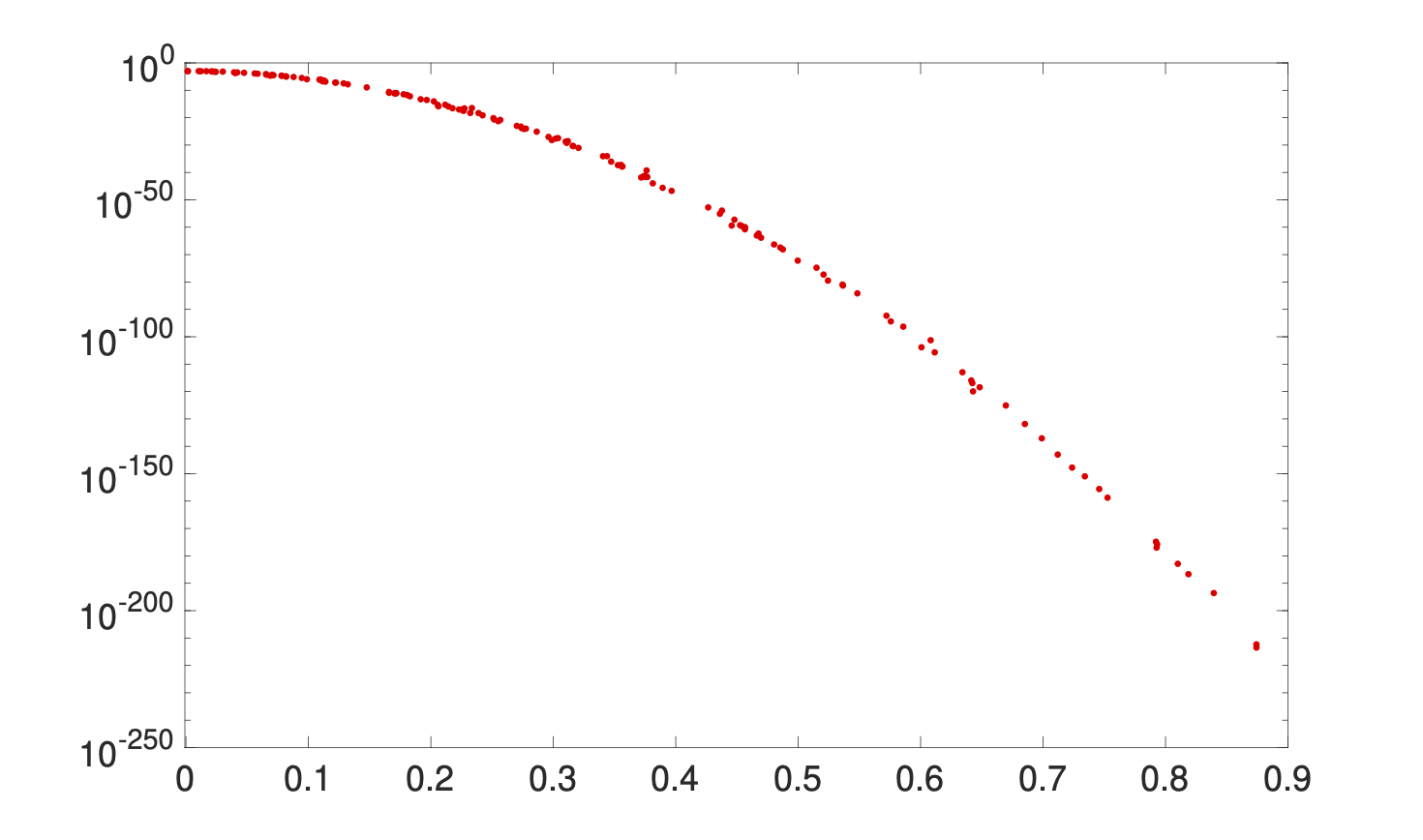} 
\caption{Weights $w_i$ assigned to the $150$ outliers from case (ii), plotted against the distance of the corresponding points $Q_i$ from the Franke function in the MEWLS approximation, for scaling factors $r = 2$ (left) and $r = 500$ (right). \label{fig_franke4}}
\end{figure}

In Figure \ref{fig_franke4}, we compare the weights $w_i$ associated with the outlier points $Q_i$ (from (ii)) against their distance from the Franke function, for the two scaling factors in \eqref{scal_fac}. Both plots indicate that, for a given reduction factor $r$, the greater the distance of a point $Q_i$, the more its corresponding weight $w_i$ is reduced. Furthermore, as expected, a higher value of $r$ causes the weights of the more distant outliers to become negligible, effectively preventing them from influencing the approximation.

To assess the ability of the entropy-based approach to automatically reduce the influence of outliers, we use the B-spline surface from item (a) above, evaluated at $\Omega$, as a reference for determining the accuracy of the model in a cross-validation procedure.  The corresponding mean squared error $\MSEcv$ decreases from $3.56 \cdot 10^{-3}$ for the OLS approximation applied to the entire dataset (Figure \ref{fig_franke2}) to $5.87 \cdot 10^{-4}$ for the MEWLS approximation with $r = 2$, eventually reaching $1.15 \cdot 10^{-6}$ for $r = 500$ (Figure \ref{fig_franke3}).

\begin{figure}
\includegraphics[width=.5\textwidth]{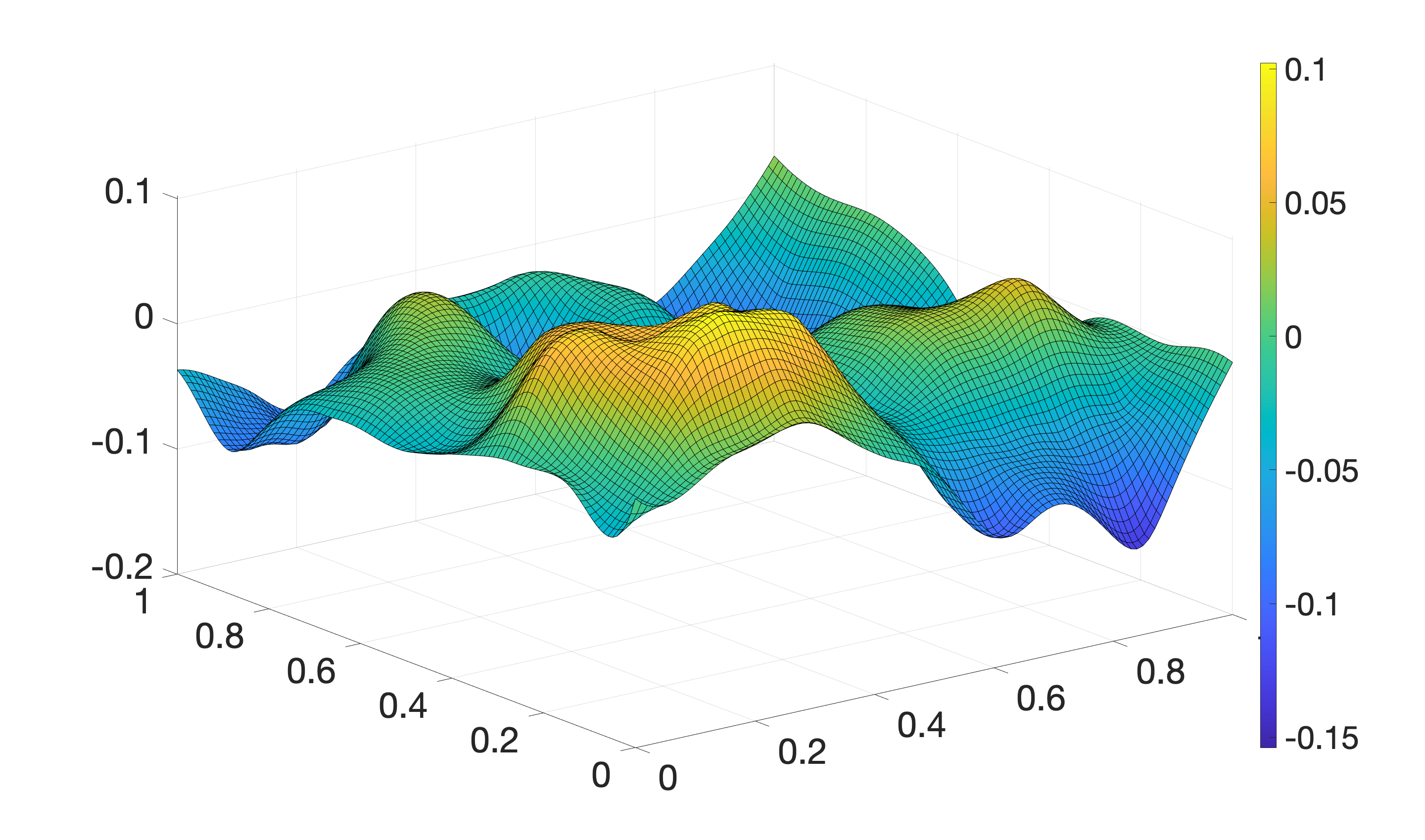} 
\includegraphics[width=.5\textwidth]{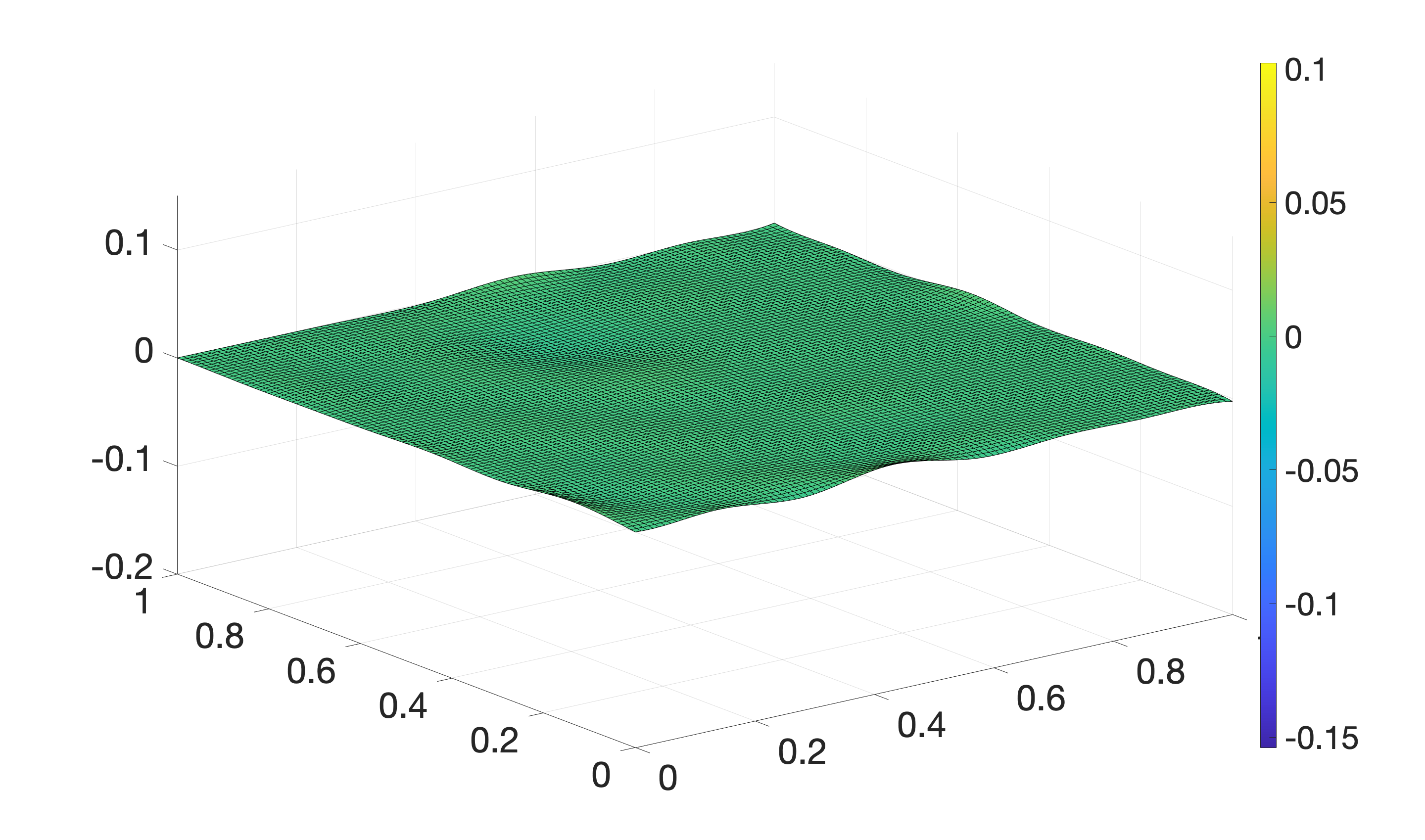} 
\caption{Error relative to the surface obtained in case (a). Left picture: thin-plate smoothing spline. Right picture: MEWLS approximation, $r=500$. \label{fig_franke5}}  
\end{figure}

Finally, for comparison, we apply the thin-plate smoothing spline using the  MATLAB function {\texttt tpaps} from  {\it Curve Fitting} toolbox, with different smoothing parameters $p$. The best performance is obtained with $p = 0.8$, yielding $\MSEcv = 2.58 \cdot 10^{-3}$, which is three orders of magnitude higher than the $\MSEcv$ achieved by the MEWLS approximation with $r=500$.  Figure \ref{fig_franke5} displays the error functions for both methods, namely the difference between the surfaces they produce and the reference surface from case (a).  

Table~\ref{tab:comparisons} provides a concise summary of the approximation performance, measured in terms of $\MSEcv$, for the three considered methods under the two scenarios previously described. In the outlier-free setting, where the data are affected only by low-variance Gaussian noise, the standard OLS approximation already yields a very accurate result. Concerning the MEWLS method, it is worth to note that the quantity $s^*$ defined in (\ref{s*}) is  essentially zero ($s^* = -7.13 \cdot 10^{-6}$), indicating that enforcing a reduction in $\MSE$ may even lead to a slight deterioration. The method \texttt{tpaps}, when finely tuned (with $p=0.99$), performs slightly better in this specific case.

In the presence of outliers, however, the behavior of the three methods changes significantly. The performance of OLS degrades considerably, whereas MEWLS, configured with a large scaling factor $r = 500$, remains highly accurate, with an $\MSEcv$ comparable to that observed in the clean data case. Although \texttt{tpaps}, with regularization parameter  $p = 0.8$, slightly outperforms OLS, its approximation error remains substantially higher than that achieved by MEWLS.

\begin{table}[h!]
\centering
\begin{tabular}{r|lll}
   & \multicolumn{1}{c}{OLS} & \multicolumn{1}{c}{MEWLS} & \multicolumn{1}{c}{tpaps}\\
\hline
absence of outliers  &  $2.37 \cdot 10^{-5}$ & $2.37 \cdot 10^{-5} ~ (r=1)  $ & $6.26 \cdot 10^{-6}$ ~ (p=0.99) \\
presence of outliers  & $3.61 \cdot 10^{-3}$ & $2.55 \cdot 10^{-5} ~ (r=500)$ & $2.58 \cdot 10^{-3}$ ~ (p=0.8)\\
\hline
\end{tabular}
\caption{Comparison of the approximation error (measured in terms of $\MSEcv$) obtained by the different methods in the absence and presence of outliers.}
\label{tab:comparisons}
\end{table}

\begin{figure}
\includegraphics[width=.5\textwidth]{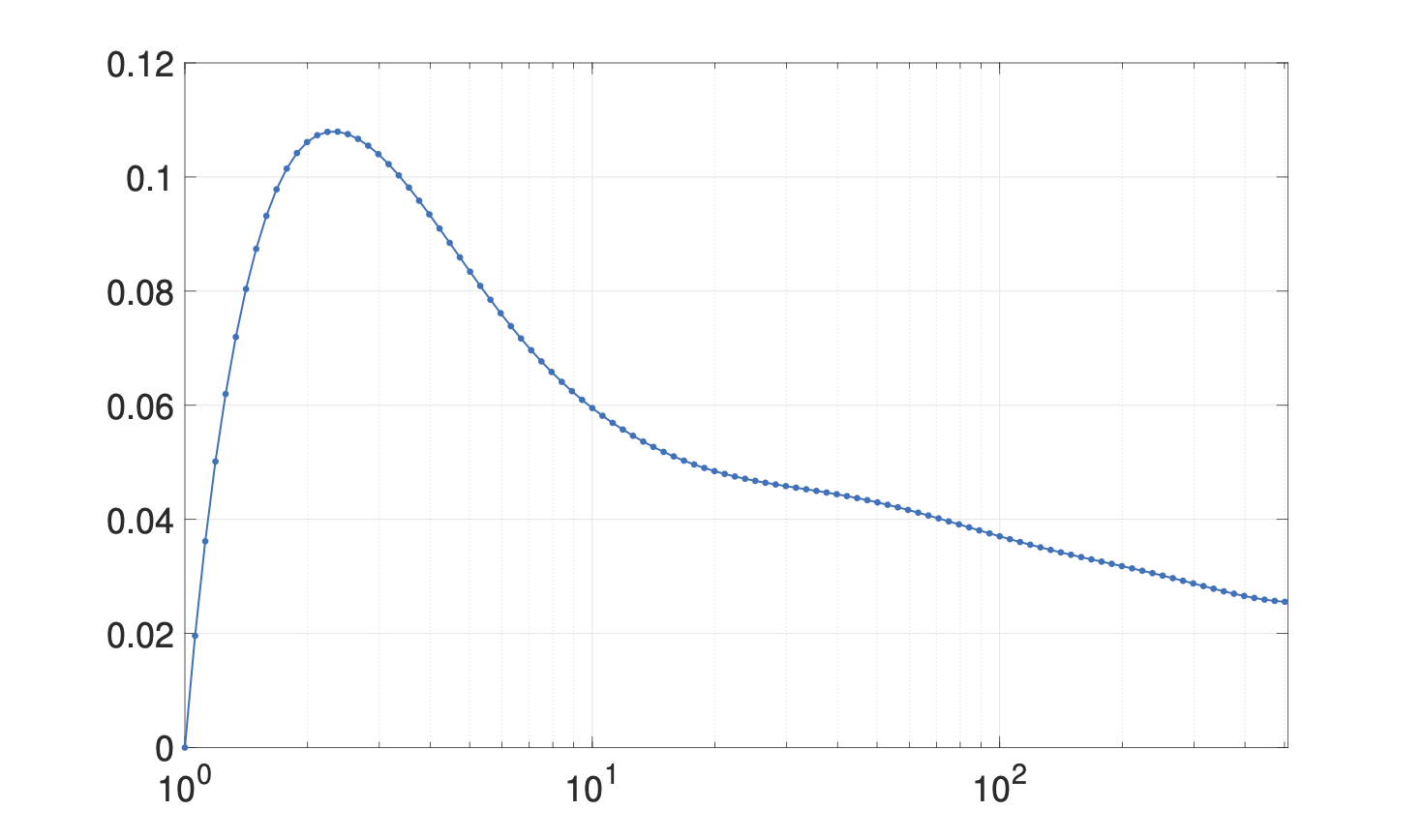} 
\includegraphics[width=.5\textwidth]{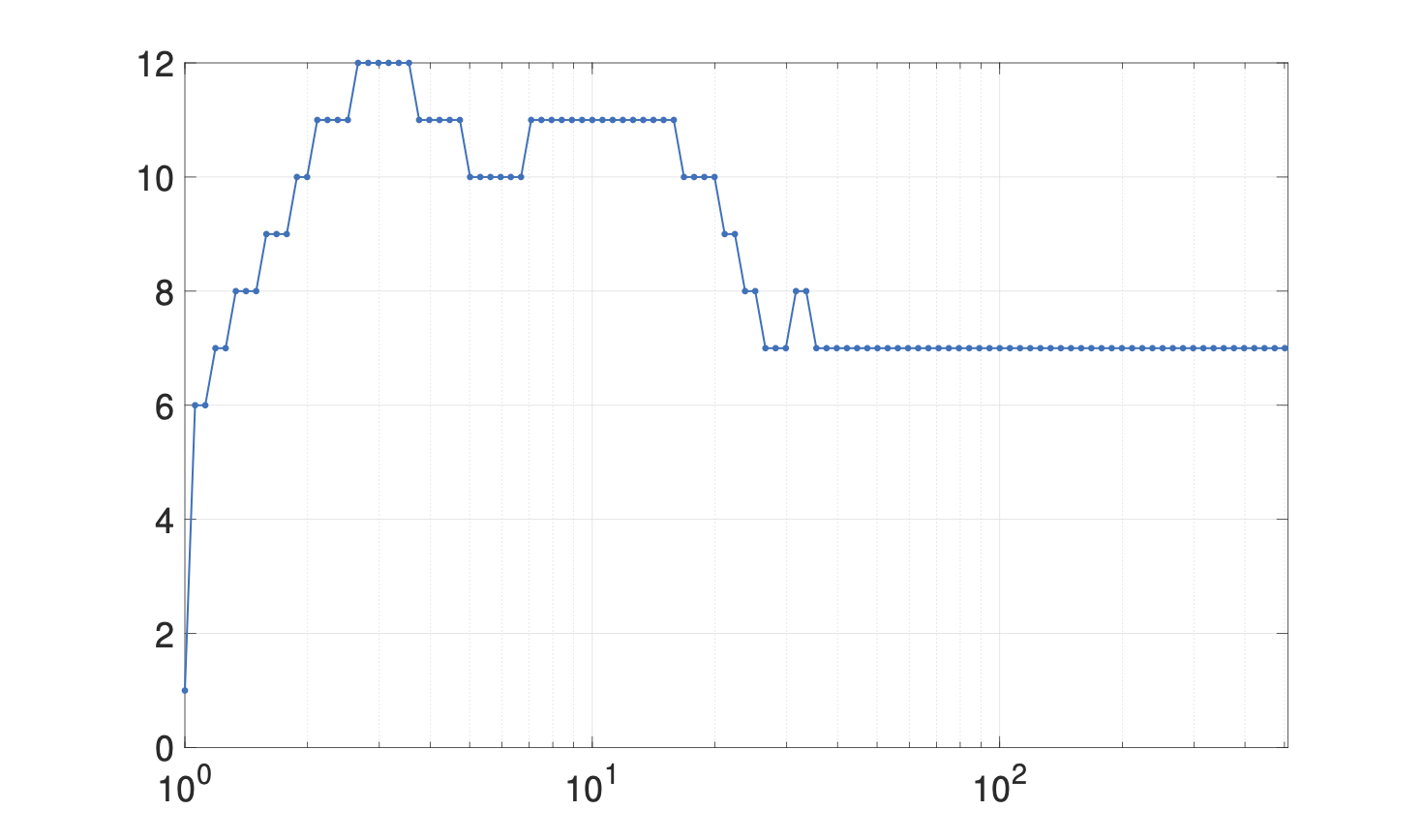} 
\caption{Behavior of nonlinear Gauss--Seidel iteration with respect to the scaling factor $r$.  Left picture: spectral radius of $G_{33}$. Right picture: number of Gauss--Seidel iterations required to attain convergence within a tolerance of $10^{-8}$. \label{fig_franke6}}  
\end{figure}

Before concluding this experiment, we briefly discuss the convergence behavior of the nonlinear Gauss--Seidel iteration employed in Algorithm \ref{Alg1}, by computing the spectral radius of the matrix $G_{33}$ defined at (\ref{G33}). In the left picture of Figure \ref{fig_franke6}, we plot the spectral radius of $G_{33}$ as a function of the scaling factor $r = \MSEuw/\MSE$ over the range $[1, 500]$. As predicted by Theorem \ref{maintheo}, the spectral radius tends to zero as $r \to 1$, indicating that the  iteration converges rapidly. Moreover, we observe that the spectral radius remains bounded and below $0.12$ throughout the considered range, thus ensuring convergence of the nonlinear Gauss--Seidel iteration even for large values of $r$. The right picture of Figure \ref{fig_franke6} displays the number of iterations required to reach convergence (a tolerance of $10^{-8}$ has been used). This behavior qualitatively mirrors that of the spectral radius: the number of iterations increases with $r$ initially, reaches a plateau, and then slowly decreases. Overall, this confirms that the Gauss--Seidel procedure is efficient and maintains fast convergence across a broad range of values of $r$.

\subsection{Reconstructing a spherical surface from noisy structured data}
In this example, we consider a $15 \times 12$ matrix of points $Q_{ij}$ wrapping around the surface of a sphere of radius 1. Each column (associated with $u_i$) represents points on a level curve, which forms a closed circle since the last point always coincides with the first. Similarly, each row (associated with $v_j$) represents a meridian curve, forming a semicircle.  

Consequently, the points in the first and last columns are identical, as both circumferences collapse to a single point. Likewise, the first and last rows are identical.  

\begin{figure}
\begin{tabular} {c}
\includegraphics[width=.5\textwidth]{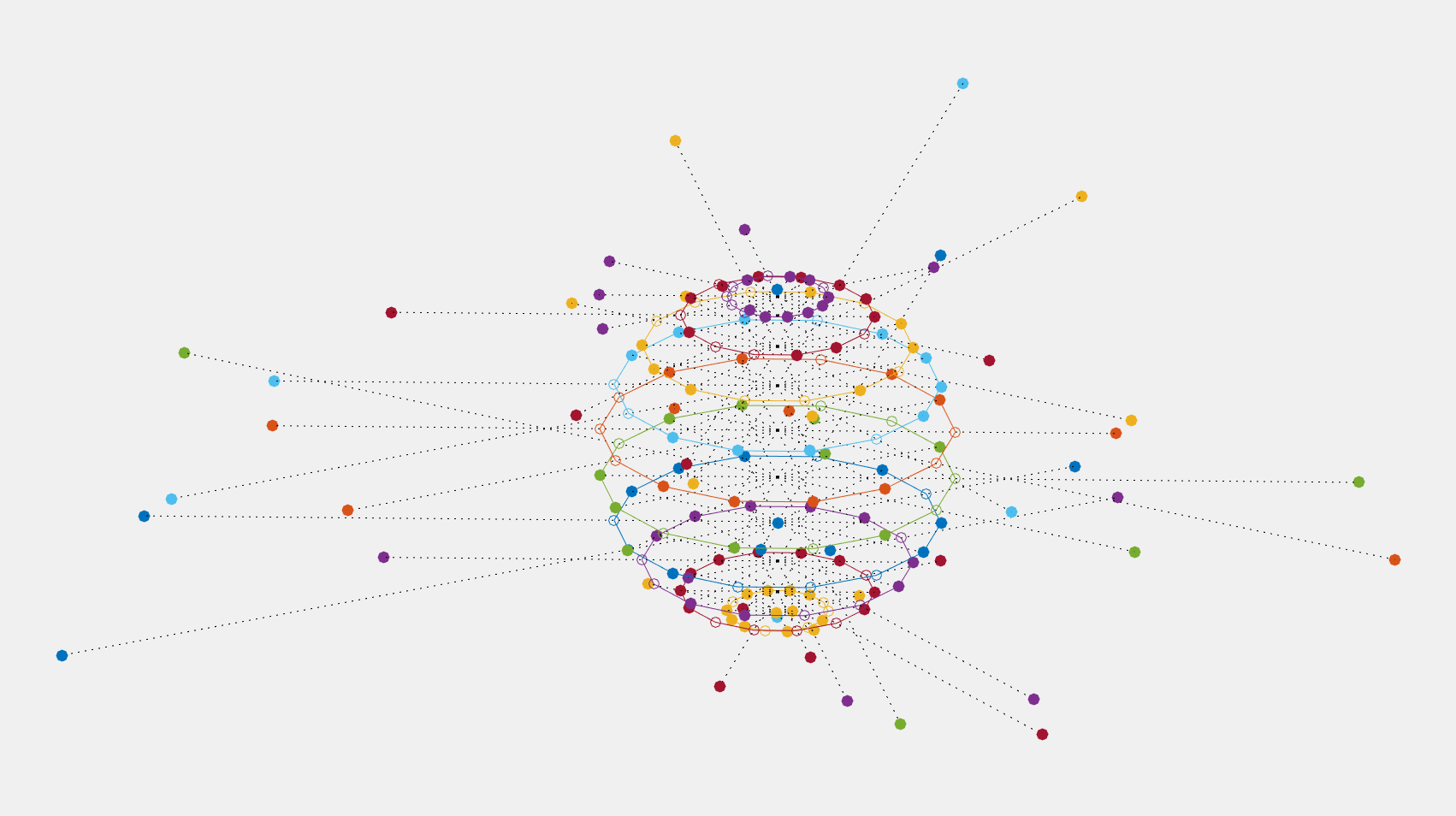} \\ \small{(a) - dataset}
\end{tabular}
\begin{tabular} {c}
\includegraphics[width=.5\textwidth]{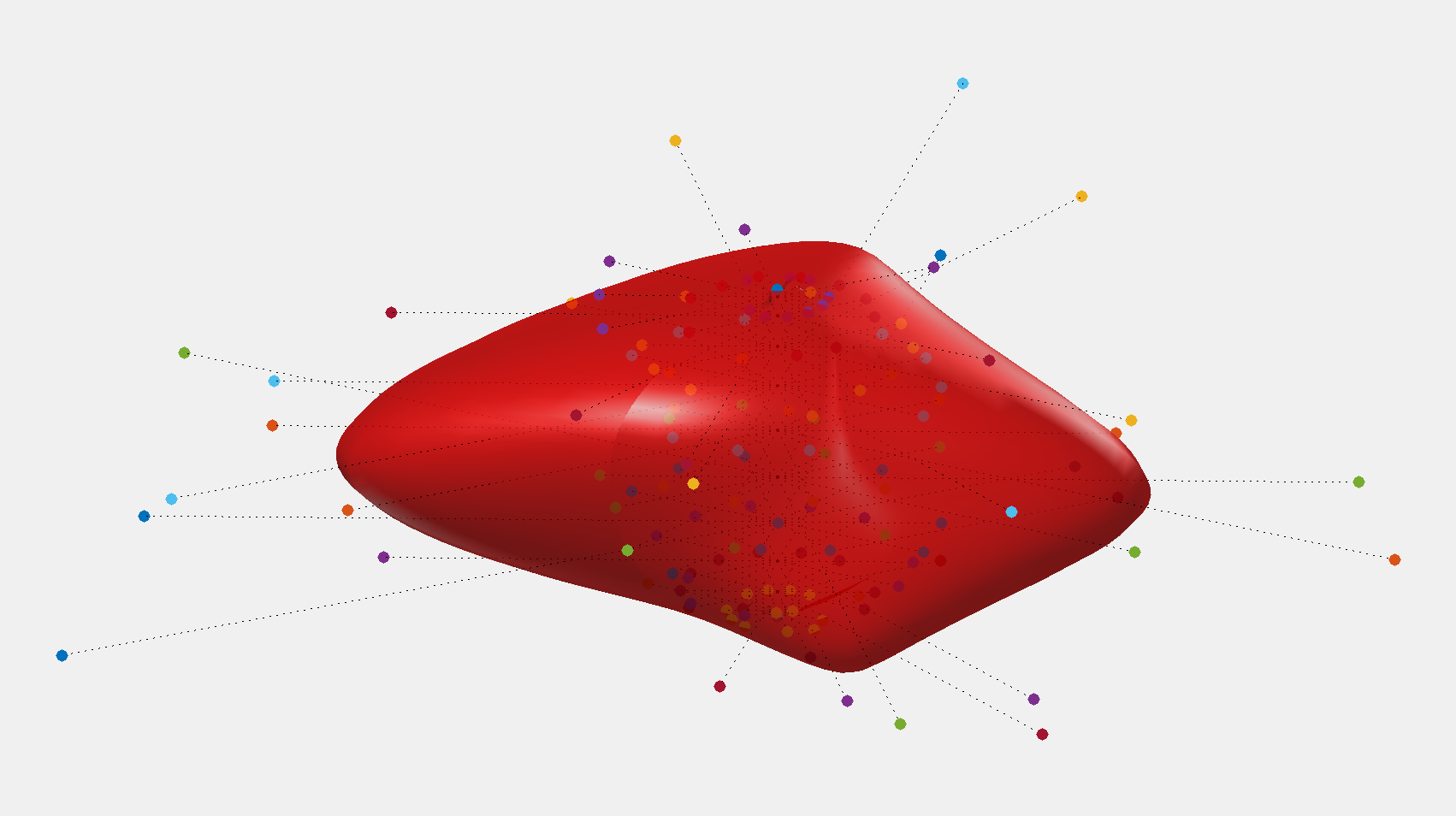} \\ \small{(b) - OLS approximation}
\end{tabular}
\begin{tabular} {c}
\includegraphics[width=.5\textwidth]{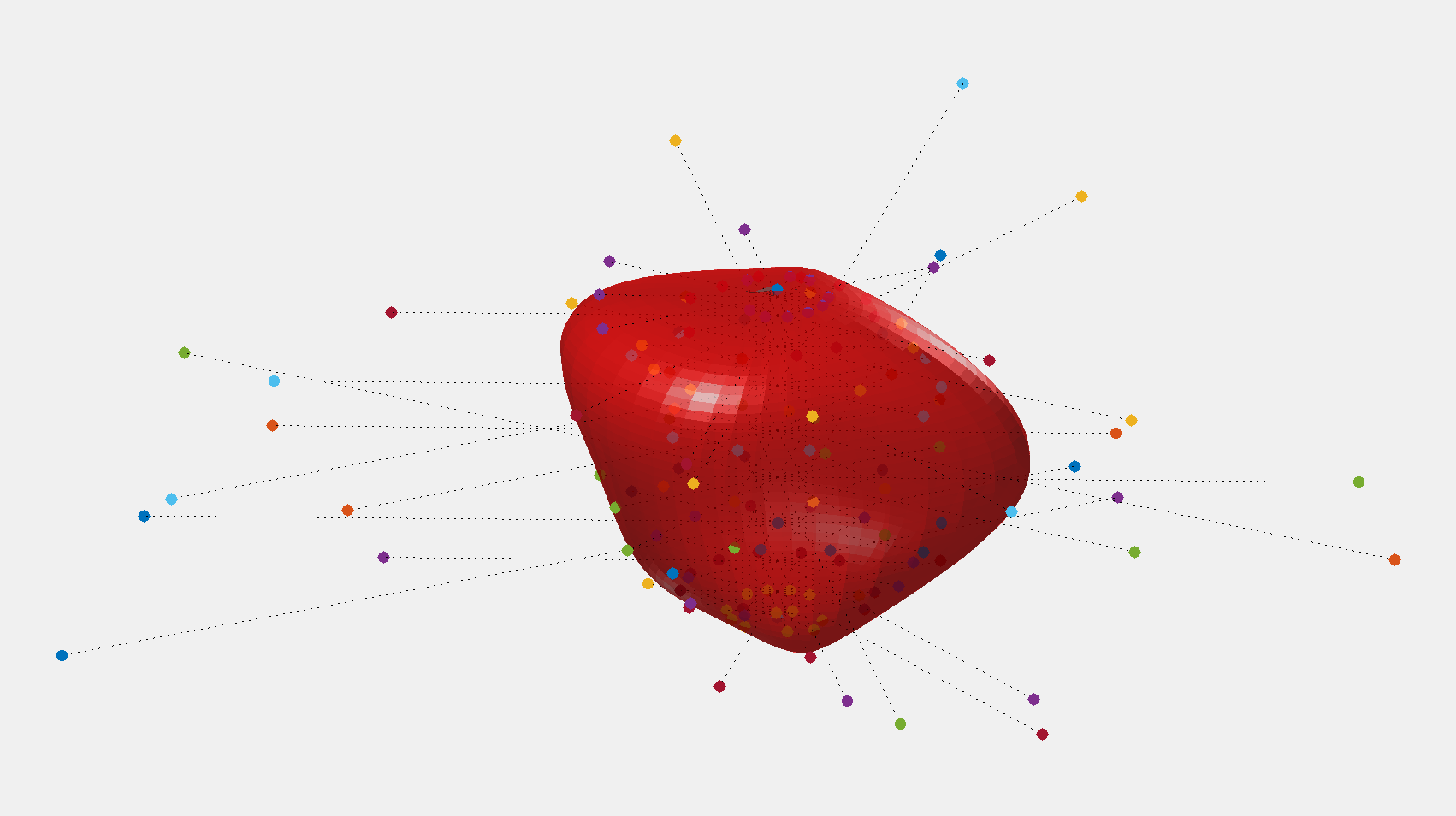} \\ \small{(c) - MEWLS approximation, $r=5$}
\end{tabular}
\begin{tabular} {c}
\includegraphics[width=.5\textwidth]{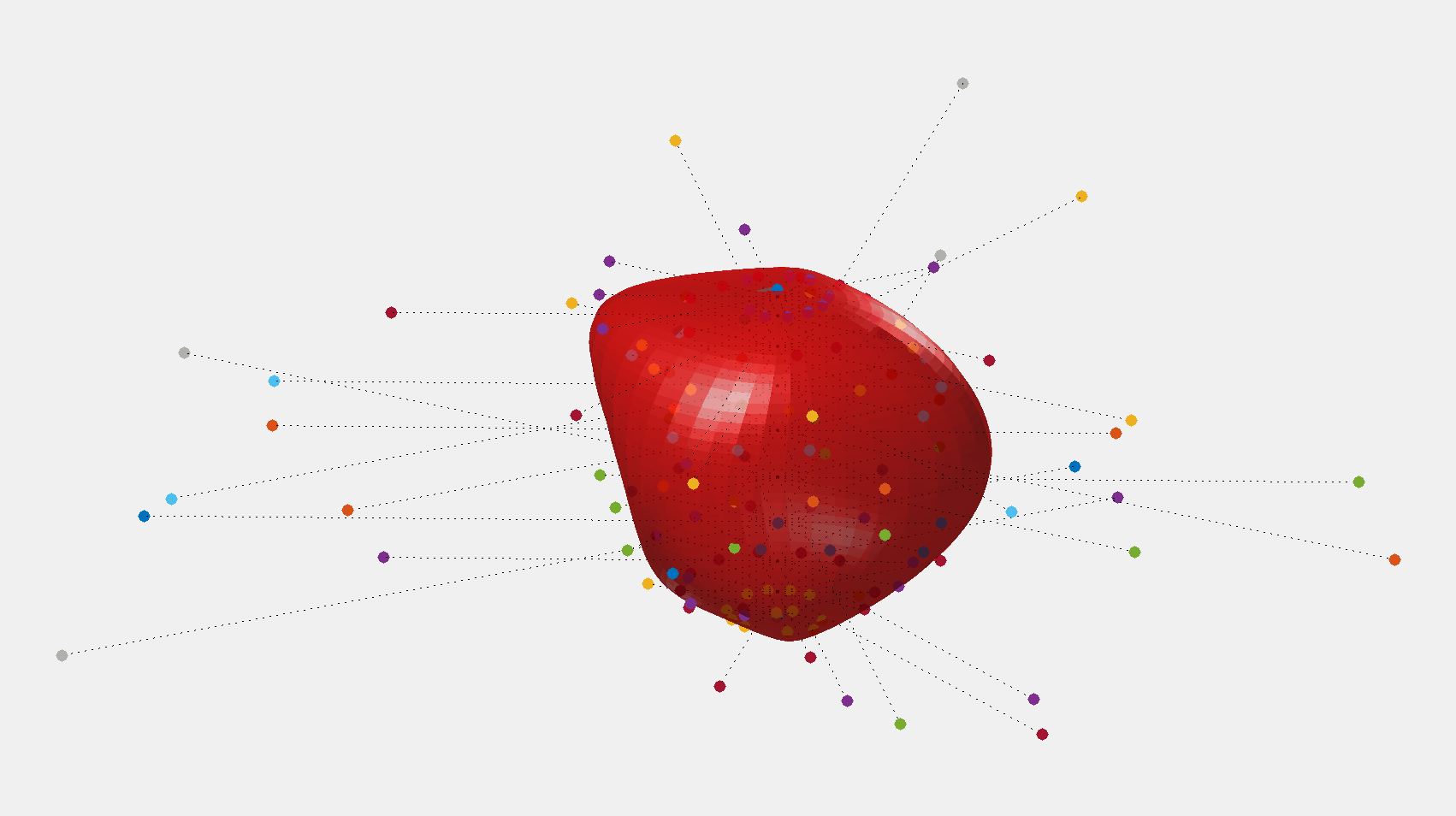} \\ \small{(d) - MEWLS approximation, $r=10$}
\end{tabular}
\begin{tabular} {c}
\includegraphics[width=.5\textwidth]{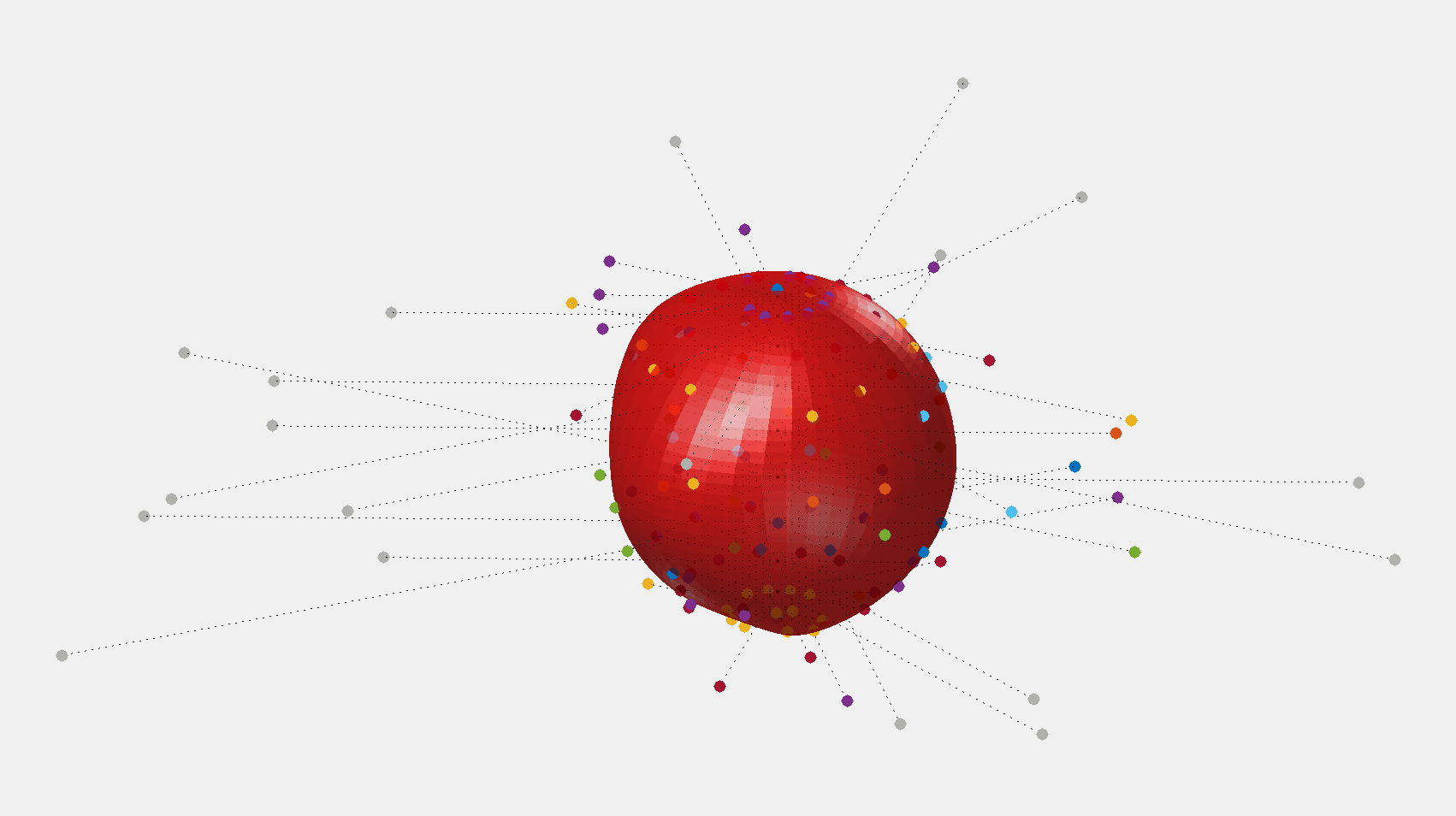} \\ \small{(e) - MEWLS approximation, $r=50$}
\end{tabular}
\begin{tabular} {c}
\includegraphics[width=.5\textwidth]{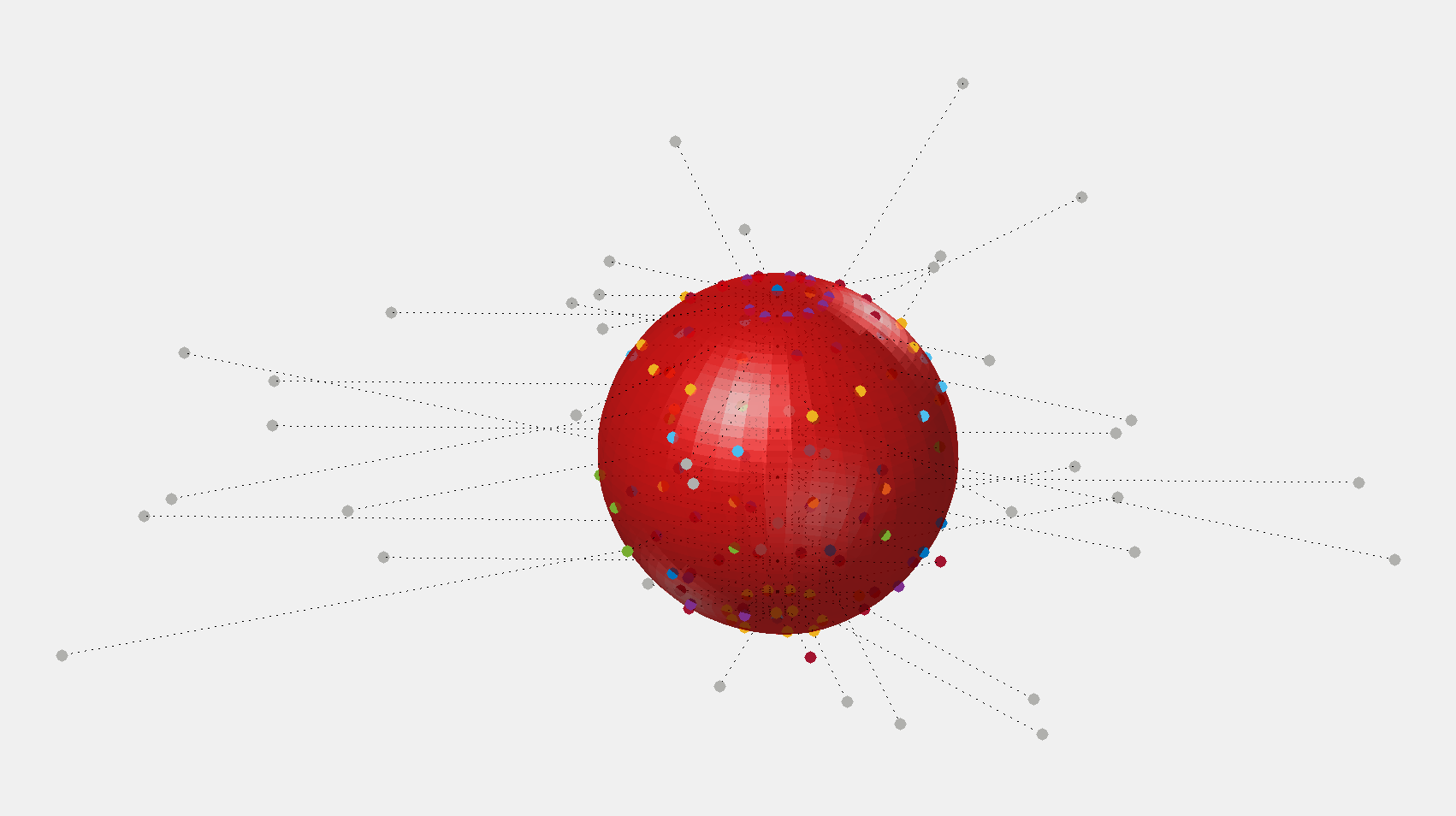} \\ \small{(f) - MEWLS approximation, $r=1000$}
\end{tabular}
\caption{Reconstruction of a spherical surface from noisy structured data. \label{fig_sphere}}
\end{figure}
Almost half of these points are perturbed, with their distance from the center varying up to four times the correct value (see Figure \ref{fig_sphere}a). The goal is to construct a closed and regular degree-3 B-spline surface on a $6 \times 5$ grid that, through the maximization of the weight distribution entropy, aims to eliminate perturbed data while preserving the spherical surface. 

For this reason, even though the unknown control points $P_{ij}$ are $30$, we construct the B-spline surface using a $9 \times 5$ grid, where the last three rows are set equal to the first three, and the 13 knots $u_i$ are chosen to be equally spaced. This ensures that the degree-3 B-spline curves $C(u)$ computed along each column satisfy $C^{(i)}(u_4) = C^{(i)}(u_{10})$ for $i = 0,1,2$, meaning they are both closed and regular.  Similarly, the nine knots $v_j$ are defined by setting the first four equal to 0, the last four equal to 1, and $v_5 = 0.5$, ensuring that each curve associated with a row of the grid interpolates at the endpoints.  Figure \ref{fig_sphere}b shows the OLS surface, where the disturbance caused by the perturbed points is clearly visible.  

Next, we consider an iterative process based on solving (\ref{constr}) for decreasing values of $\MSE$. Starting with a uniform weight distribution and the corresponding value of $\MSEuw$, we compute the MEWLS approximation by setting $\MSE = \MSEuw / r$, progressively increasing the reduction factor $r$ from $5$ to $1000$.  We can observe that the weights associated with the perturbed values gradually decrease and eventually become negligible compared to the others. Figures \ref{fig_sphere}c-e show intermediate results for $r = 5, 10, 50$. A point $Q_{ij}$ is considered negligible and is therefore represented by an empty (gray) dot when its corresponding weight satisfies $w_{ij} < 10^{-7}$. 

Figure \ref{fig_sphere}f shows the final result, obtained with $r = 1000$. In this case, about 70\% of the perturbed points become negligible (in particular, the more distant ones), while all correct points receive nearly identical, non-negligible weights, ensuring they contribute equally to the final approximation. 

\section{Applications to real-world datasets}
\label{sec_applications}
In this section, we evaluate the practical applicability of our method on two real-world problems, illustrating its effectiveness beyond synthetic datasets. The first application focuses on craquelure detection and reduction in oil paintings, where the goal is to identify fine crack patterns and minimize their visual impact while preserving the integrity of the artwork. The second application addresses tumor detection in mammograms, where our method is used to enhance the identification of dense masses in breast tissue. These examples highlight the robustness of our approach in handling noisy and irregular data while maintaining essential structural features.

\subsection{Removing craquelure from an oil painting}
Craquelure refers to the intricate network of fine cracks that gradually emerge on the surface of oil paintings, especially those executed on wooden panels. This phenomenon arises naturally over time and is due to the aging of materials and environmental influences. The primary cause is the different expansion and contraction rates between the paint layers and the wooden support in response to fluctuations in temperature and humidity, leading to the formation of these characteristic fissures. 

\begin{figure}
\begin{center}
\begin{minipage}{.2\textwidth}
\begin{center}
original image  \\
$101575$ pixels
\end{center}
\end{minipage}
\begin{minipage}{.75\textwidth}
\includegraphics[width=0.9\textwidth]{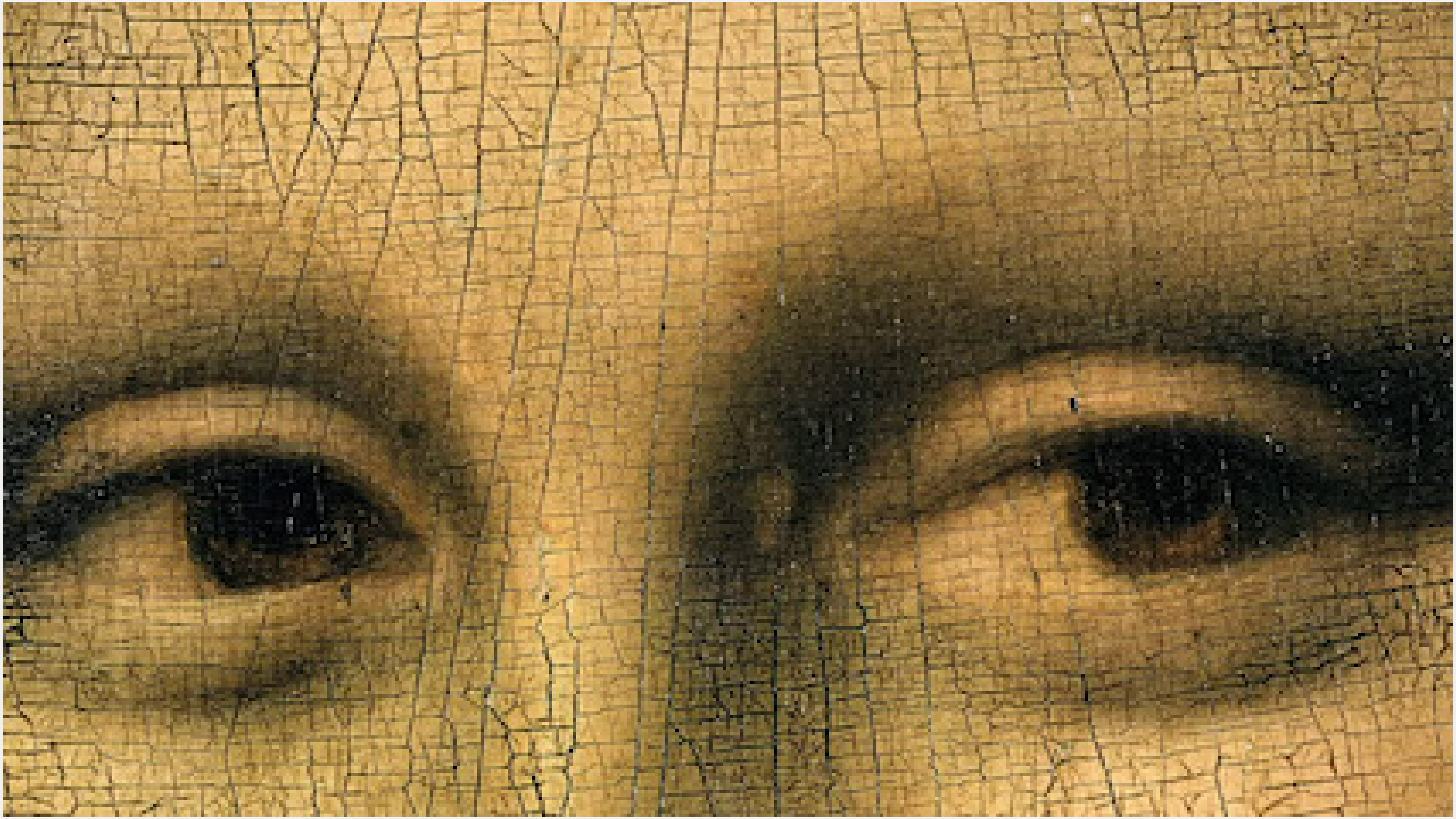}  \\[.2cm]
\end{minipage}
\begin{minipage}{.2\textwidth}
\begin{center}
detected outliers \\
13951 pixels
\end{center}  
\end{minipage}
\begin{minipage}{.75\textwidth}
\includegraphics[width=0.9\textwidth]{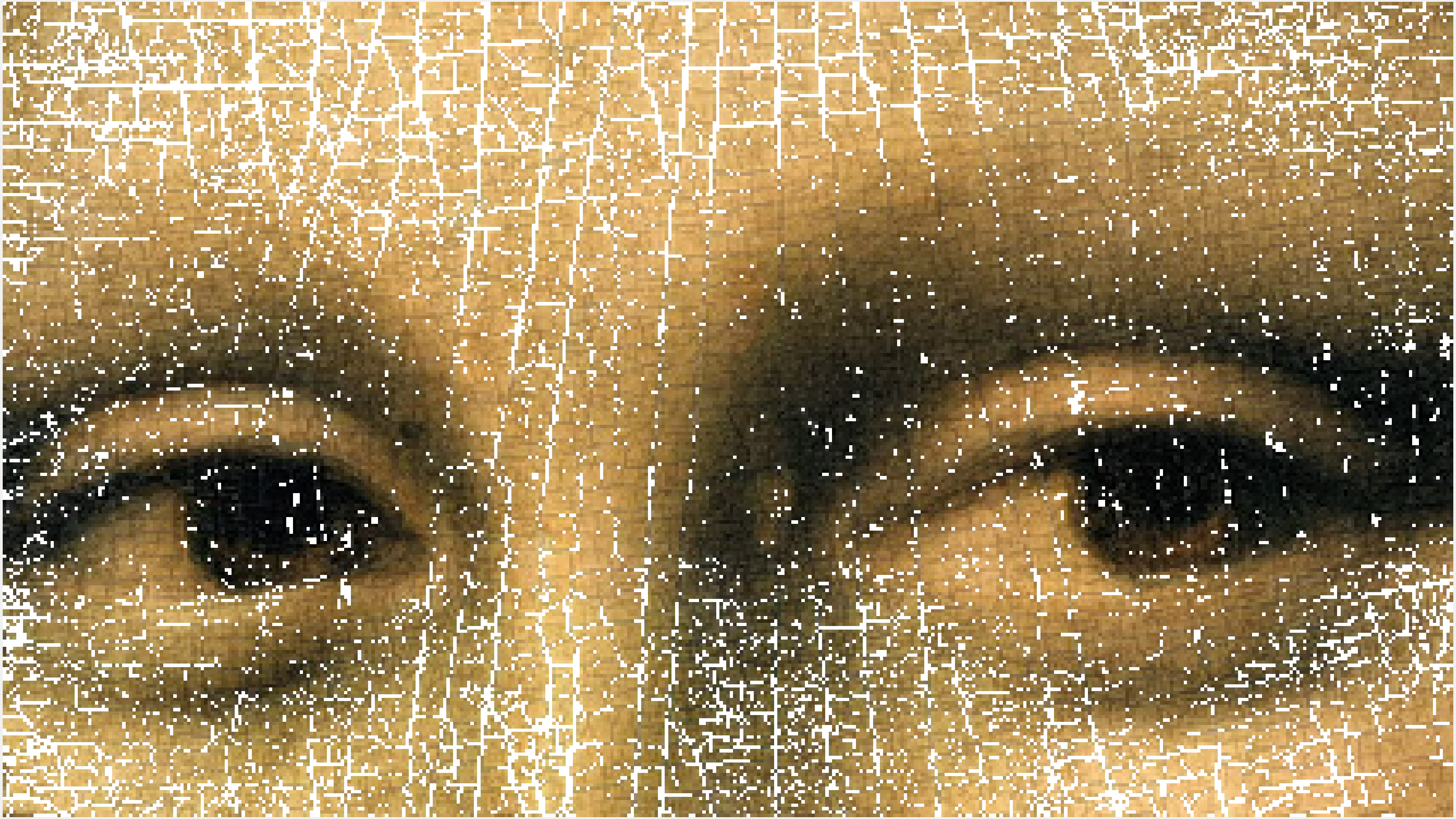} \\[.2cm]
\end{minipage}
\begin{minipage}{.2\textwidth}
\begin{center}
restored image  \\
$13.7\%$ pixels replaced
\end{center}
\end{minipage}
\begin{minipage}{.75\textwidth}
\includegraphics[width=0.9\textwidth]{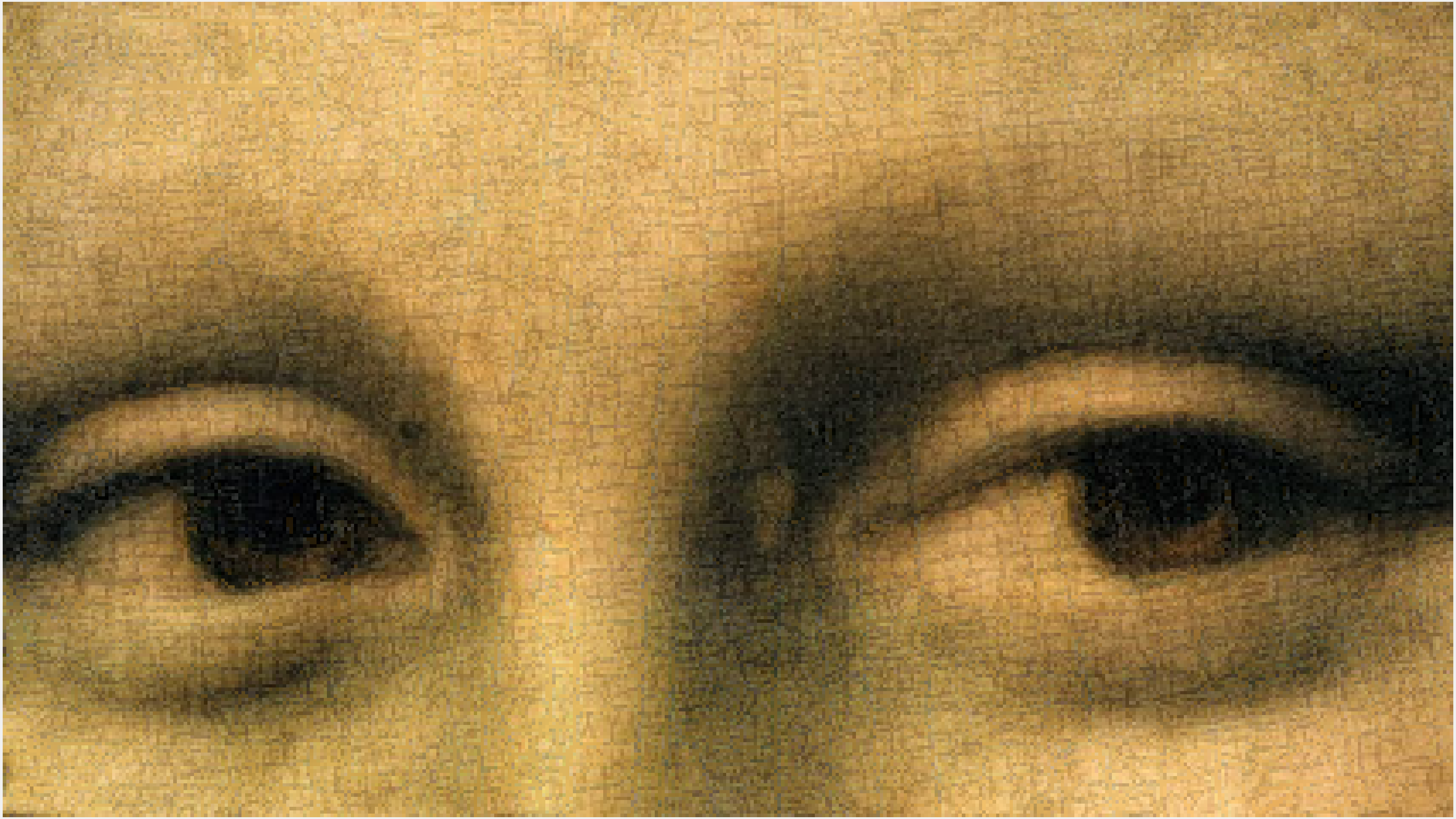} 
\end{minipage}
\caption{Top picture: original image portraying Mona Lisa's eyes. Middle picture: pixels detected as outliers by Algorithm \ref{Alg1} are highlighted in white. Bottom picture: restored image.}
\label{fig:monnalisa}
\end{center}
\end{figure}

One of the most famous examples is Leonardo da Vinci's Mona Lisa, where the fine pattern of cracks is particularly visible on her face. The top image of Figure \ref{fig:monnalisa} shows the area surrounding her eyes. Our aim here is to apply the maximum entropy approach to detect most of the pixels forming the pattern of cracks and minimize their visual impact while limiting pixel-level intervention.

To this end, we solve the constrained problem (\ref{constr}) by considering a bivariate spline $S(u,v)$ defined on a uniform grid of dimension $40 \times 60$ and selecting $\MSE=\MSEuw/2$ as input value in the second constraint. From the array of weights produced by Algorithm \ref{Alg1}, a binary mask $\Omega$ is generated to identify the weights that the procedure has reduced to sufficiently small values, according to the following definition 
$$
\Omega_{ij} = \left\{ \begin{array}{ll}
1, & \mbox{if } w_{ij} < w_{\mbox{\scriptsize max}}/10, \\
0, & \mbox{otherwise}.
\end{array}
\right.
$$

It turns out that the nonzero elements of the matrix $\Omega$ amount to $13951$ out of the $239 \times 425 = 101575$ pixels that compose the original image. The middle image of Figure \ref{fig:monnalisa} shows the resulting mask $\Omega$ as white pixels superimposed on the original image. A comparison with the top picture highlights the high level of accuracy achieved. 

Finally, we replace the original RGB values of the detected outliers with those predicted by the model. The resulting image, shown at the bottom of Figure \ref{fig:monnalisa}, retains $86.3\%$ of the pixels from the original source image, while only $13.7\%$ of the pixels required restoration.

\subsection{Detecting masses in mammograms}
\begin{figure}
\begin{center}
\includegraphics[width=.4\textwidth]{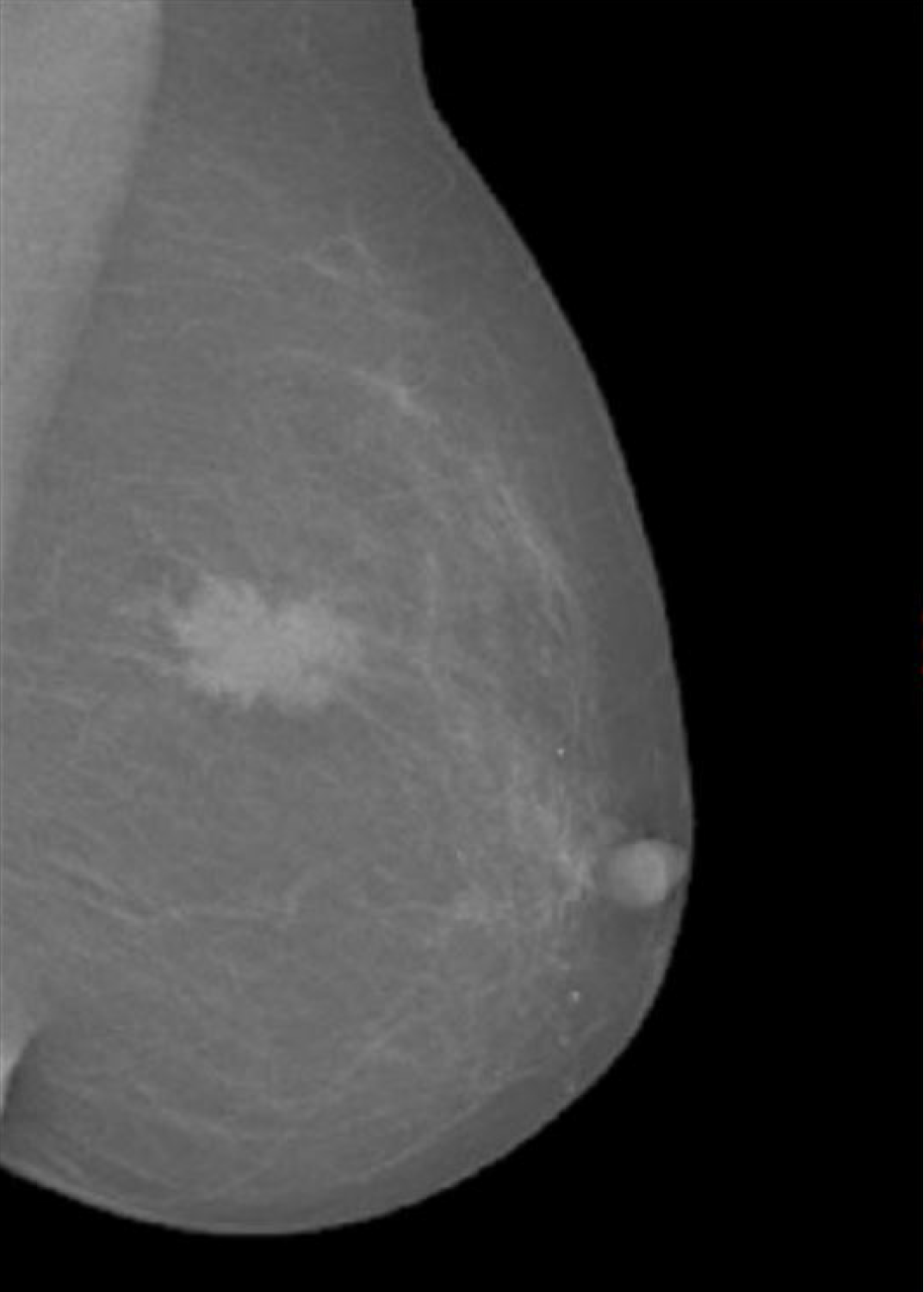}  \hspace*{.5cm}
\includegraphics[width=.4\textwidth]{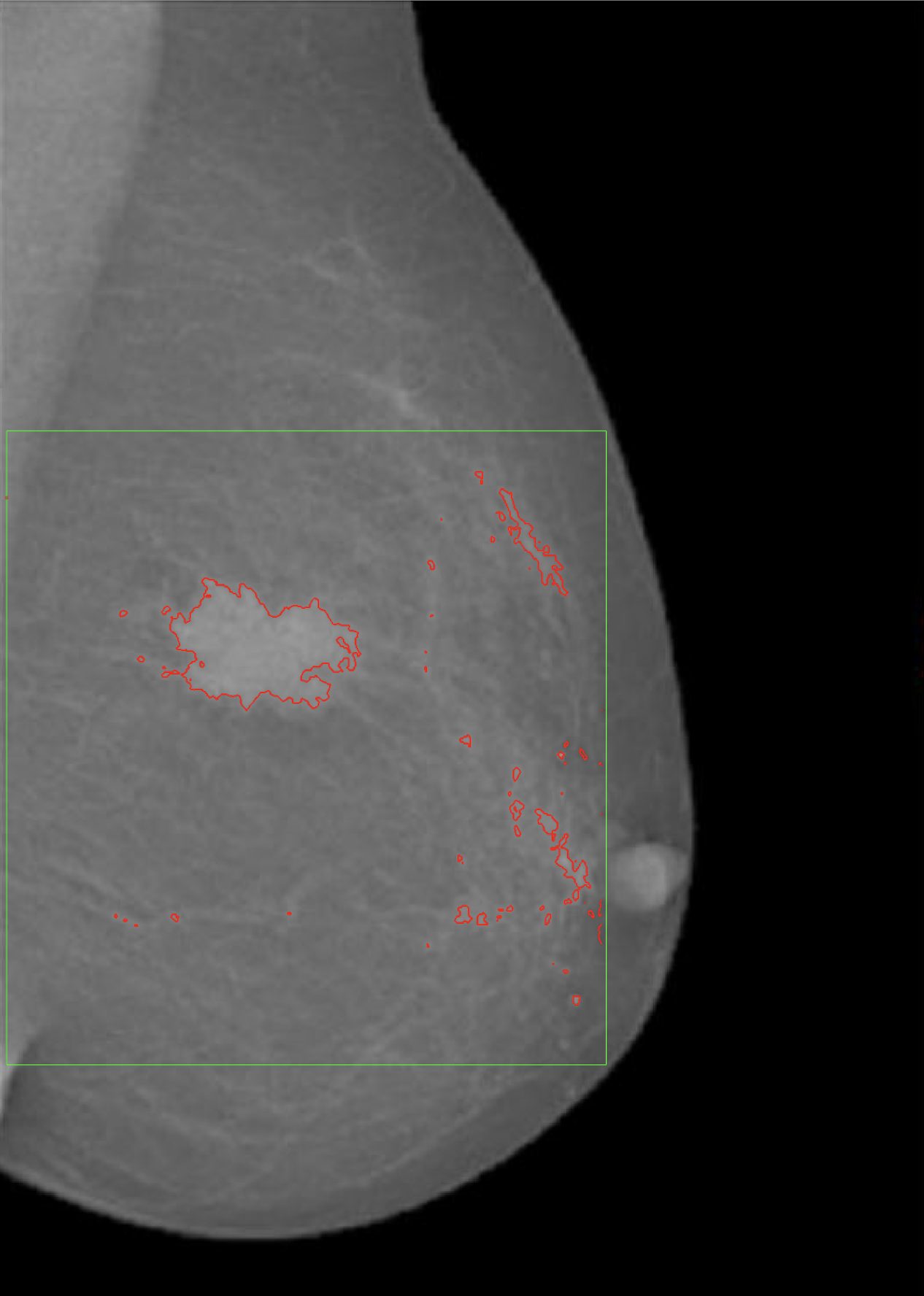} \\
\caption{Left picture: a mammogram revealing a dense mass in the central region. Right picture: Regions of Interest are highlighted as red contours superimposed to the original source image.}
\label{fig:breast-1}
\end{center}
\end{figure}

As a proof of concept, we have used the MEWLS approach as a computer-aided diagnosis tool to identify Regions of Interest (ROIs) and assist radiologists in making accurate diagnoses of breast lesions. The mammogram shown in the left picture of Figure \ref{fig:breast-1} reveals a dense mass (a malignant tumor) with an irregular and blurry boundary in the central region \cite{MDM22}.

\begin{figure}
\begin{center}
\includegraphics[width=.45\textwidth]{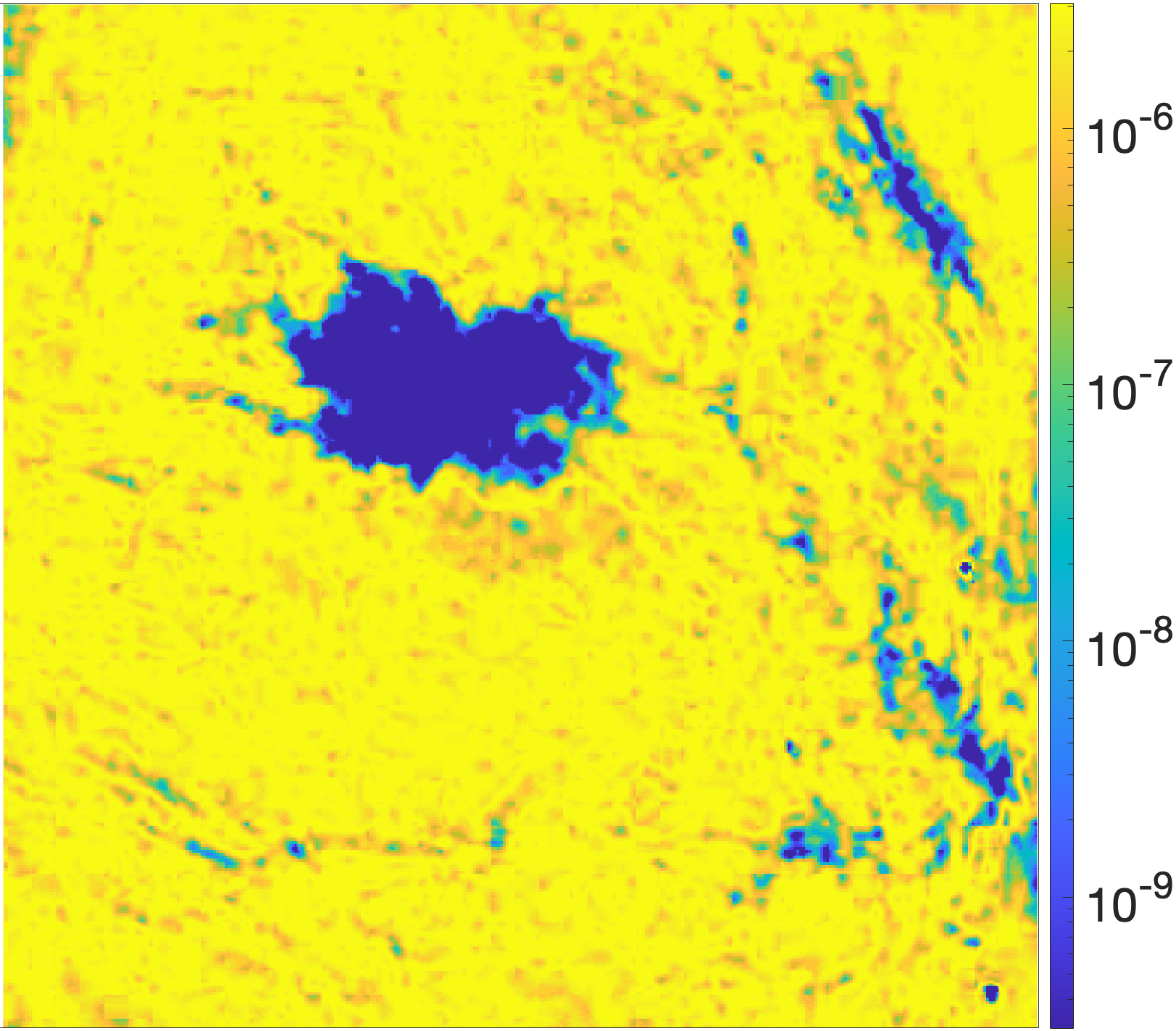}  \hspace*{.5cm}
\includegraphics[width=.45\textwidth, height=.3\textheight]{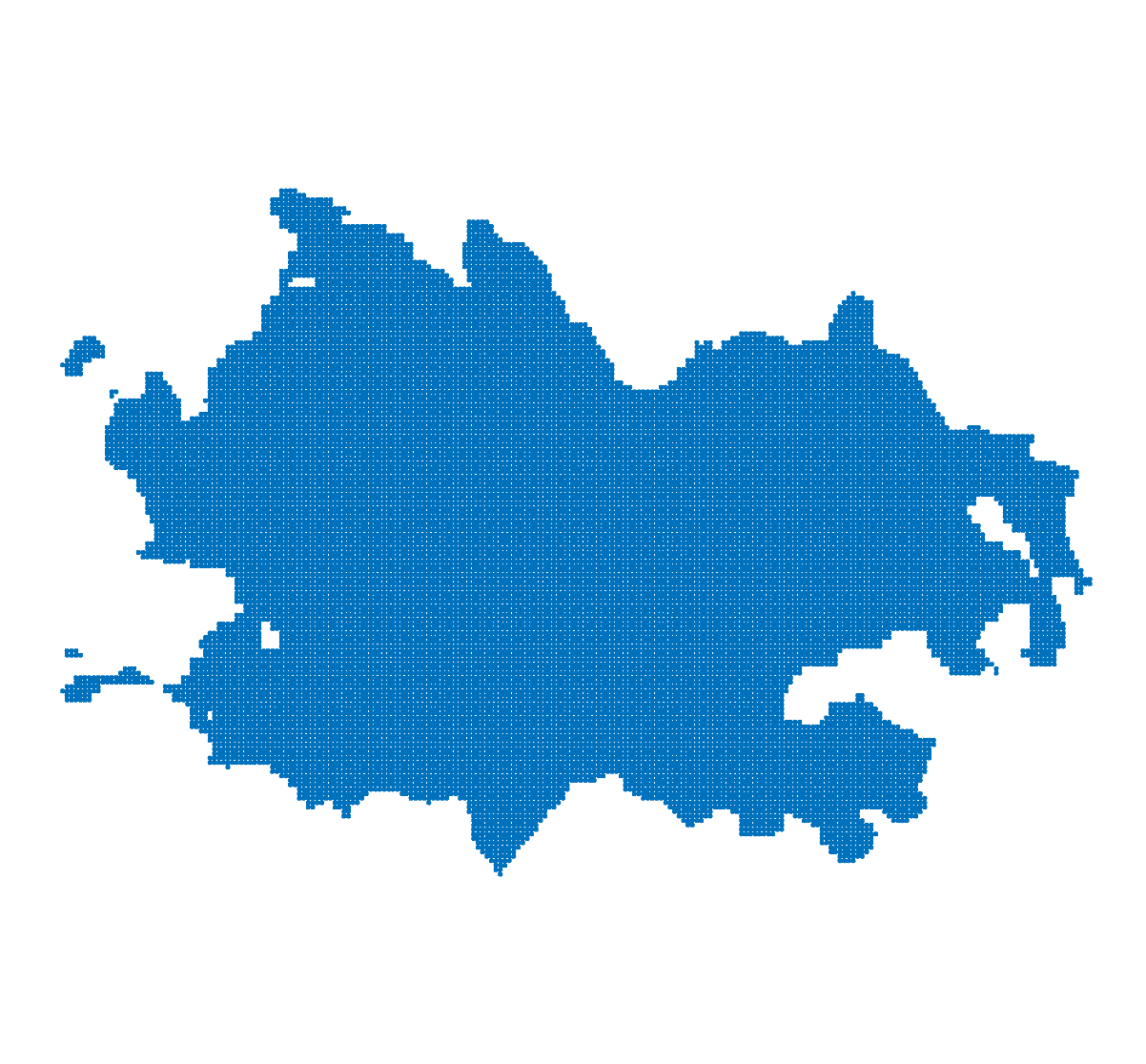}
\caption{Left picture: weight distribution obtained from the MEWLS approximation. Right picture: binary mask highlighting the contour of the dense mass.}
\label{fig:breast-2}
\end{center}
\end{figure}

We have solved problem (\ref{constr}) using a cubic B-spline defined on a $6 \times 6$ grid covering a rectangular domain that embraces most of the breast area (green line in the right picture of Figure \ref{fig:breast-1}). The original $\MSEuw$ value, corresponding to uniform weights $w_i\simeq 2.11\cdot 10^{-6}$, has been reduced by a factor of $5$ to obtain the weight distribution shown in the left picture of Figure \ref{fig:breast-2}. Here, lighter shades correspond to higher weight values, whereas darker shades correspond to smaller weights.

We observe that the MEWLS approximation assigns very small weights to pixels corresponding to the interior of dense masses. This allows us to extract, from the weight distribution image,  level sets associated with the most significant intensity variations, thereby identifying the boundaries of the ROIs, which are superimposed on the original mammogram in the right picture of Figure \ref{fig:breast-1}.

In \cite{RaNg07}, the authors explore the use of fractal dimension as a feature for classifying breast masses in mammograms based on their contour complexity. Their study is based on the observation that malignant masses tend to have more irregular, spiculated boundaries, whereas benign masses typically exhibit smoother, more regular contours. Their analysis, performed on a dataset of $57$ masses—$37$ benign  and $20$ malignant—reveals that the average fractal dimension of the $37$ benign masses is $1.13 \pm 0.05$ (mean $\pm$ standard deviation) while  the average fractal dimension of the $20$ malignant tumors is $1.41 \pm 0.15$. 

Motivated by this study, as an additional investigation into the nature of the tumor mass in Figure \ref{fig:breast-1}, we compute the Hausdorff fractal dimension of its boundary. To achieve this, we extract a binary image from the weight distribution (left picture of Figure \ref{fig:breast-2}), where pixels corresponding to the mass are labeled as $1$, while background pixels are labeled as $0$. This binary representation is displayed in the right picture of Figure \ref{fig:breast-1}.

Using the Matlab function {\em hausDim}, available in \cite{Co25}, we obtain a Hausdorff fractal dimension of the mass boundary equal to $1.59$. This result aligns with the findings in \cite{RaNg07} and raises suspicion that the mass may be malignant which, in this case, is indeed confirmed.

\section{Conclusions}
\label{sec_conclusions}
By formulating the problem within a weighted least-squares framework, we have introduced a robust approach for bivariate spline approximation that employs the principle of maximum entropy to optimize weights and mitigate the influence of outliers. Maximizing the entropy of the weight distribution ensures that the probabilities are spread as evenly as possible, subject to the constraint that the Mean Squared Error (MSE) reaches a prescribed value. As a result, this diffusion process assigns smaller weights to anomalous data points while preserving the underlying structure of the dataset.

Our synthetic and real-world experiments confirm the effectiveness of the method in reconstructing surfaces from noisy and outlier-contaminated datasets, demonstrating its practical relevance and adaptability to structured data.

While promising, the method involves certain aspects that warrant further exploration. In particular, it requires the specification of the target $\MSE$ parameter, which influences the balance between fidelity to the data and robustness against outliers. Moreover, the entropy optimization step entails solving constrained nonlinear systems, which may become computationally demanding for large datasets. Addressing these challenges could involve developing more efficient large-scale algorithms and adaptive, data-driven approaches for selecting the $\MSE$ parameter. A first attempt in this direction has been illustrated in \cite{DeFaIaLoMaRu25}.

Given its general framework, several avenues for future research remain open. One direction is extending the approach to adaptive knot placement, further enhancing accuracy and computational efficiency. Another promising direction is its integration into nonlinear model fitting problems, particularly those arising in machine learning and deep learning applications.

\section*{Acknowledgements}  We wish to  thank INdAM-GNCS for its valuable support, under the INdAM-GNCS Project   CUP\_E53C24001950001. P. Amodio and F. Iavernaro acknowledge the contribution of the National Recovery and Resilience Plan, Mission 4 Component 2 - Investment 1.3 - Future Artificial Intelligence Research - FAIR - PE00000013 - Spoke 6 - Symbiotic AI,  under the NRRP MUR program funded by the European Union - NextGenerationEU - (CUP\_H97G22000210007). We also wish to thank ``{\em Fondo acquisto e manutenzione attrezzature per la ricerca}'' D.R. 3191/2022, Universit\`a degli Studi di Bari Aldo Moro.


\end{document}